\documentclass[12pt]{elsarticle}
\usepackage{amsmath}
\usepackage{amssymb}
\usepackage[figuresright]{rotating}

\journal{Journal of \LaTeX\ Templates}

\begin{document}

\begin{frontmatter}

\title{Design and Calibration of the 34 GHz Yale Microwave Cavity Experiment}





\author[YalePhysics]{P. L. Slocum}
\author[YalePhysics]{O. K. Baker}
\author[YalePhysics,OmegaP]{J. L. Hirshfield}
\author[YalePhysics]{Y. Jiang}
\author[YalePhysics]{A. T. Malagon}
\author[YalePhysics]{A. J. Martin}
\author[YalePhysics]{S. Shchelkunov}
\author[YalePhysics,YaleAstro]{A. Szymkowiak}

\address[YalePhysics]{Department of Physics, Yale University, PO Box 208120, New Haven, CT USA 06520}
\address[OmegaP]{Omega-P, Inc., 291 Whitney Ave.,Suite 401, New Haven, CT 06511}
\address[YaleAstro]{Department of Astronomy, Yale University, PO Box 208101, New Haven CT  06520}

\begin{abstract}
Several proposed models of the cold dark matter in the universe
include light neutral bosons with sub--eV masses.  In many cases
their detection hinges on their infrequent interactions with
Standard Model photons at sub--eV energies.  We describe the design
and performance of an experiment to search for aberrations from the
broadband noise power associated with a 5~K copper resonant cavity
in the vicinity of 34~GHz (0.1~meV).  The cavity, microwave
receiver, and data reduction are described.  Several configurations
of the experiment are discussed in terms of their impact on the
sensitivity of the search for axion--like particles and hidden
sector photons.
\end{abstract}


\end{frontmatter}


\section{Introduction}
\label{sec:introduction}

Direct detection of cold dark matter (CDM) is required before many
questions can be addressed about its origin and implications in the
universe.  Cosmological evidence for the existence of dark
matter~\cite{Zwicky:1937zza, Sofue:2008wt, Jarosik:2010iu,
Ade:2013ktc} has consistently driven the need for measurements that
can impose additional constraints on theory.

Searches for CDM with masses of GeV/c$^2$ are ongoing but have thus
far yielded either inconclusive~\cite{Aalseth:2012if,
Angloher:2011uu,Bernabei:2008yi,Savage:2008er,Agnese:2013rvf} or
negative
results~\cite{Aprile:2013doa,Angle:2011th,Armengaud:2011cy,Armengaud:2012pfa,Akimov:2011tj,Akerib:2013tjd}.
Similarly, collider searches for signals from CDM have been
negative~\cite{ATLAS:2012ky}.  These measurements have led to
constructive limits on models as well as a heightened interest in
light CDM candidates with masses less than 1~eV. Thus far searches
for light CDM in the laboratory have focused primarily on particles
that should interact infrequently, but predictably, with Standard
Model (SM) photons. The pseudoscalar
axion~\cite{Peccei:1977hh,Weinberg:1977ma,Wilczek:1977pj} is
required to couple to 2 SM photons by way of the Primakoff effect.
Other pseudoscalar and scalar axion--like particles (ALPs) that
arise in string theory~\cite{Jaeckel:2007ch,Arias:2012mb} are
allowed, but not required, to couple to 2 photons.  The vector
hidden sector photon~\cite{Jaeckel:2007ch,Arias:2012mb} can also be
the CDM~\cite{Nelson:2011sf,Arias:2012mb} and interacts with SM
photons through mass--dependent kinetic mixing.

The nearly monoenergetic distribution of light CDM, as well as its
possible interaction with sub--eV ($\lesssim$GHz) photons, make it
susceptible to discovery in searches that utilize radio frequency
(RF) detection techniques.  In particular, the approach pioneered by
Sikivie~\cite{Sikivie:1983ip,Sikivie:1985yu} and the ADMX
collaboration~\cite{Peng:2000hd, bradley, Asztalos:2009yp} in which
a resonant cavity sitting in a strong magnetic field is coupled to a
microwave receiver can be an extremely sensitive probe of the
sub--eV mass regime~\cite{Asztalos:2009yp}. In this paper we discuss
the design and calibration of a similar experiment. A low--noise
cryogenic amplifier and microwave receiver are employed to search
for deviations from the nominal spectrum of 34 GHz power associated
with a 5~K Cu resonant cavity. Several configurations of the
experiment are discussed in terms of their sensitivity to sub--eV
mass CDM models.

\section{Experiment\label{sec:experiment}}

\subsection{Magnet and Cryostat}
\begin{figure}
\hspace{-0.5in}\includegraphics[width=6.3in]{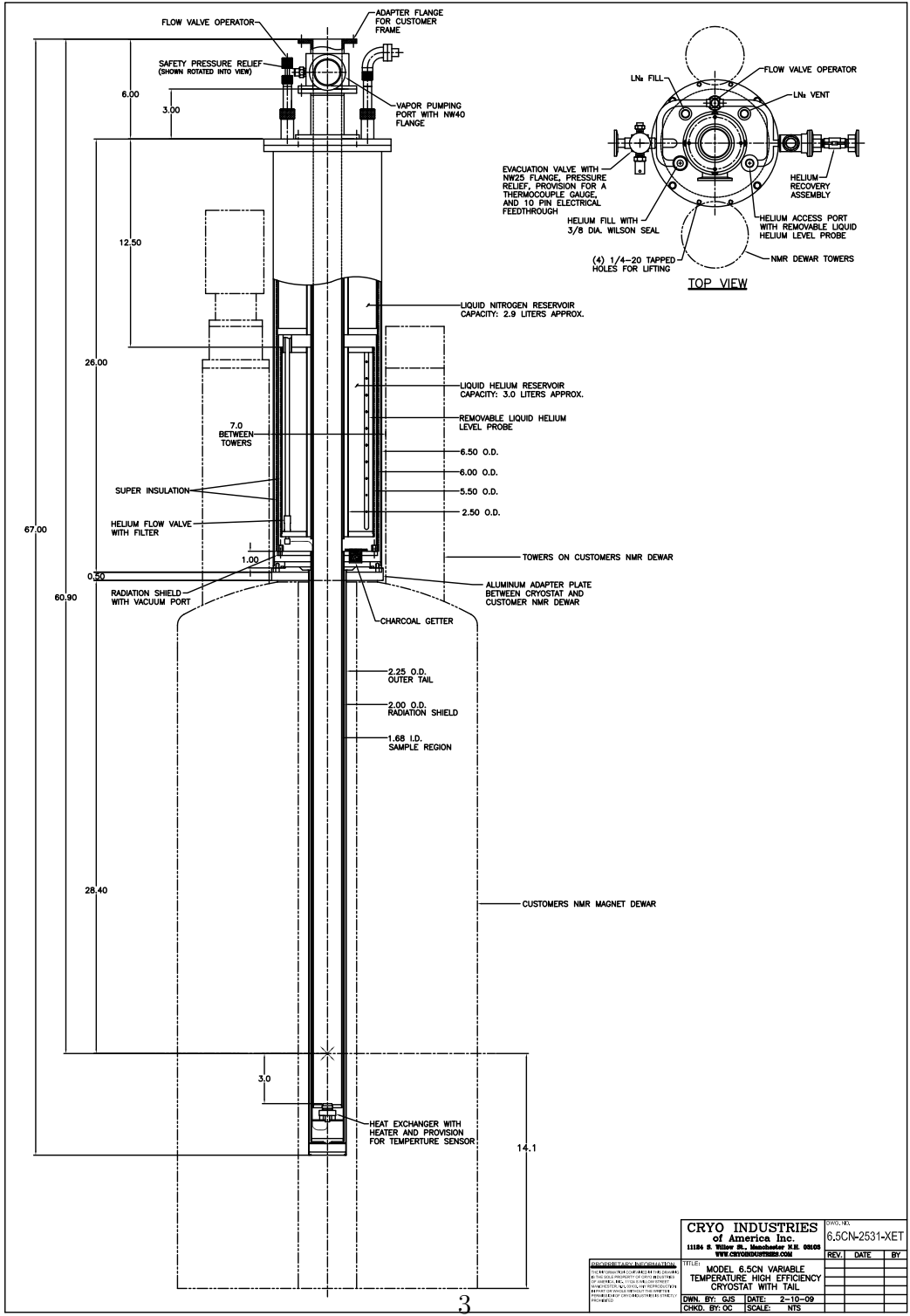}
\caption{Magnet, built by Oxford Instruments, and cold gas cryostat
built by Cryo Industries of America, Inc.\label{fig:cryostat}}
\end{figure}

The experiment sits inside an Oxford Instruments V22460
superconducting magnet with a peak field of 7~T inside the warm bore
of diameter 89~mm. The magnetic field is strongest in a 10~cm long
region located 36~cm above the bottom edge of the bore. The
intended purpose of the magnet is for measurements of nuclear
magnetic resonance (NMR) with a field that is specified to be
uniform across the peak region to a few parts per million.  In the
present experiment, however, the field uniformity is only required
to be of order 10\%.  The temporal stability of the field has been
found to be better than a few percent over a span of 5~years.  An
outline of the magnet is included in Figure~\ref{fig:cryostat}.

Also shown in Figure~\ref{fig:cryostat} is the liquid He cryostat
system that extends into the bore of the magnet.  Its central sample
tube houses the signal cavity and cryogenic amplifiers, and is
cooled by He gas that has been vaporized from a reservoir of liquid
He. Surrounding the sample tube is a liquid N$_2$ thermal radiation
shield.  The inner diameter of the sample region is 4.27~cm and is
limited by the diameter of the bore of the magnet and the size of
the room-temperature drive cavity.  The physical temperatures at the
cavity, amplifiers, and vaporizer are monitored with thin--film
resistance cryogenic temperature sensors made by Lake Shore 
Cryotronics, Inc.

\subsection{Copper Resonant Cavities}\label{sec:RunningModes}
\begin{sidewaysfigure}
\vspace{0.0in} \includegraphics[width=9.in]{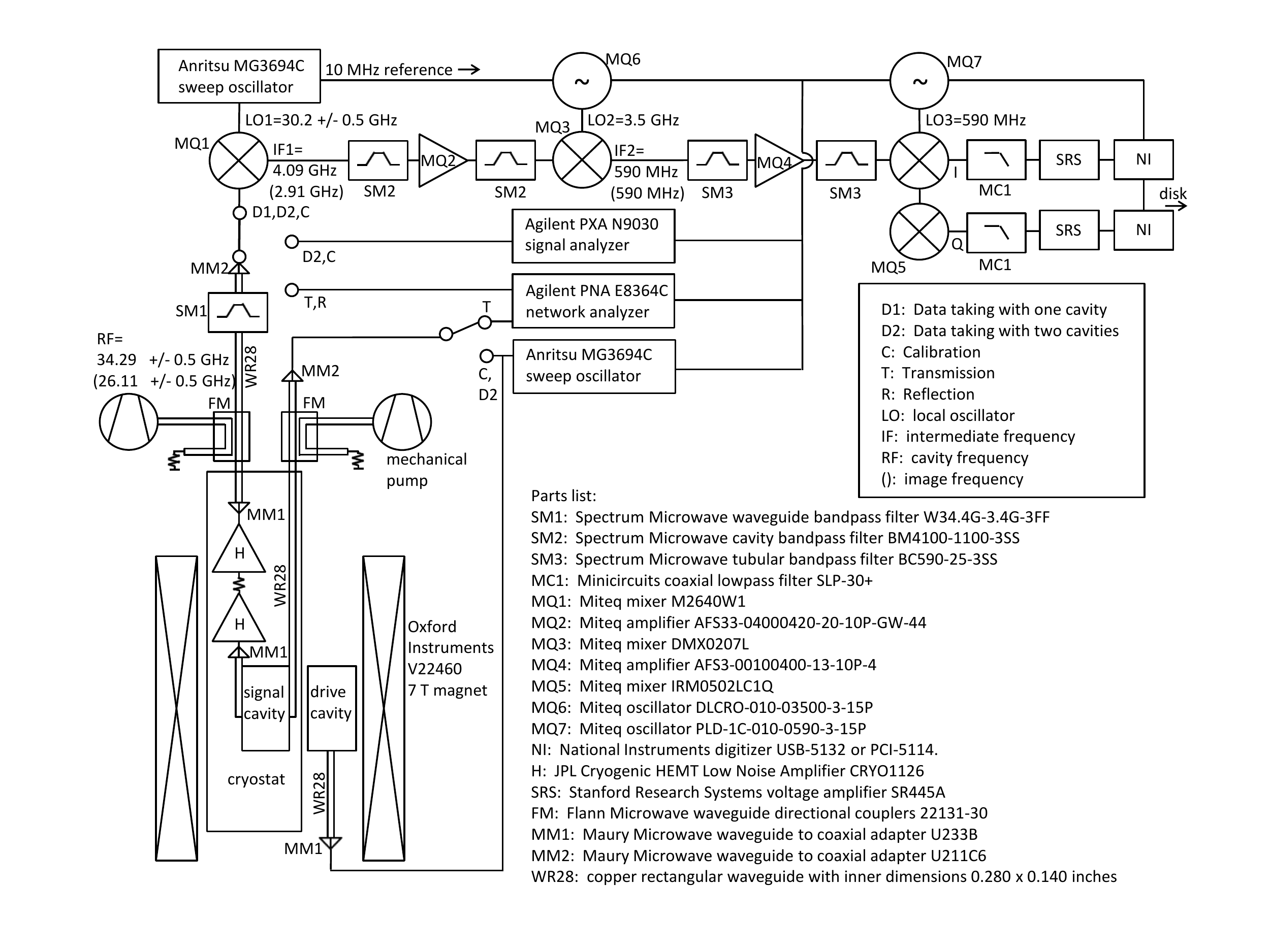}
\caption{Schematic of the experiment.\label{fig:circuit}}
\end{sidewaysfigure}

The experiment has two main running configurations:  an experiment
that is driven with 34 GHz RF power, and alternatively, a listening
mode. The driven experiment consists of two adjacent 34 GHz
oxygen--free high thermal conductivity (OFHC) copper microwave
cavities each of which supports a tranverse electric (TE) mode.  The
``signal'' cavity sits near the bottom of the cryostat in the region
where the external B-field is maximal. The ``drive'' cavity sits
adjacent to the signal cavity, inside the bore of the magnet but
outside the cryostat. In the second configuration, or listening
mode, only the signal cavity is employed.  One of two types of
signal cavities is utilized:  Either a cavity that supports a TE
mode, or one that supports the transverse magnetic (TM) mode. The
locations of the signal and drive cavities are shown in the
schematic of Figure~\ref{fig:circuit}.

\subsubsection{Signal Cavities}
\begin{figure}
\hspace{2.7in}\includegraphics[width=1.in]{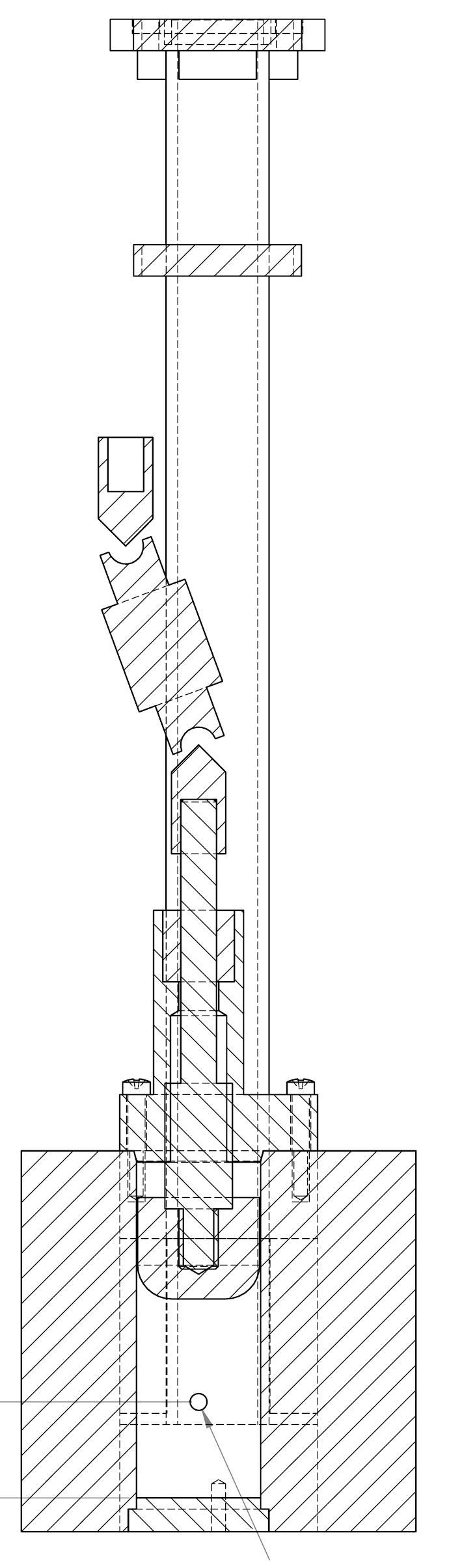}
\caption{Cross section showing the signal cavity, the plunger with
bellows for tuning, and the 2 WR28 waveguides oriented vertically.
The circle inside the outline of the cavity denotes the weakly
coupled port.  Also shown are the threaded fittings and the connector that
interfaces with the fiberglass G10 rod. \label{fig:signalcavity}}
\end{figure}

The cylindrical TE signal cavity supports the TE$_{011}$ mode and is
11~mm in diameter by 17~mm in height.  Its central resonant
frequency is 34.29~GHz. The resonant frequency is tunable across 500
MHz using a plunger that moves vertically. There is a 1~mm gap
between the plunger and the side wall whose purpose is to break the
degeneracy between the TE$_{011}$ and the TM$_{111}$ modes.  The
position of the plunger is adjusted by a lever attached to a
threaded fitting that moves freely at temperatures between 7~K and
300~K. The fitting attaches to a meter--long fiberglass G10 rod that
is turned by hand from outside the top flange of the cryostat.
Figure~\ref{fig:signalcavity} shows a drawing of the TE$_{011}$
cavity and its tuning mechanism.  Also shown are the thin OFHC fins
that increase the surface area of the assembly, helping to optimize
its rate of cooling inside the cryostat.  The fins are asymmetric
due to constraints on available space.

The TE$_{011}$ signal cavity is designed to be critically coupled at
cryogenic temperatures to a few inches of copper rectangular WR28
waveguide.  The waveguide has inside dimensions 0.280 by 0.140
inches and cutoff frequency 21.1~GHz. It terminates at a waveguide
to coax adapter and cryogenic amplifier. The power loss in the
waveguide at frequency 34~GHz is acceptable at approximately 1~dB/m.
In addition to the critically coupled port, the signal cavity is
weakly coupled to 3~m of WR28 waveguide that ends outside the
cryostat at a directional coupler. The directional coupler serves
two purposes simultaneously: It allows for vacuum pumping and it is
a port for test signals. The orientation of the WR28 waveguides
relative to the signal cavity are shown in Figure~\ref{fig:circuit}.

\begin{figure}
\hspace{-0.2in}
\includegraphics[width=3.0in]{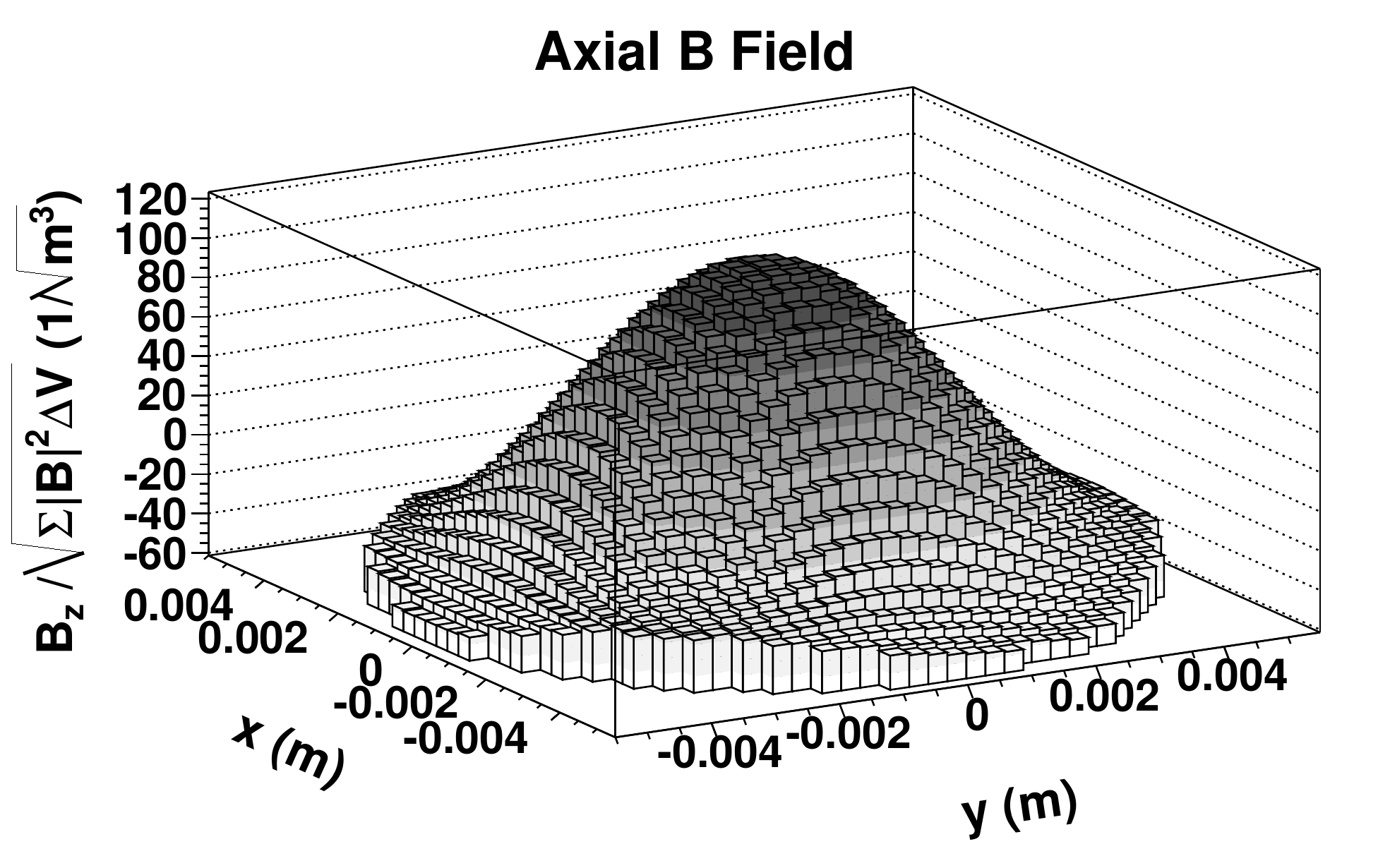}
\includegraphics[width=3.0in]{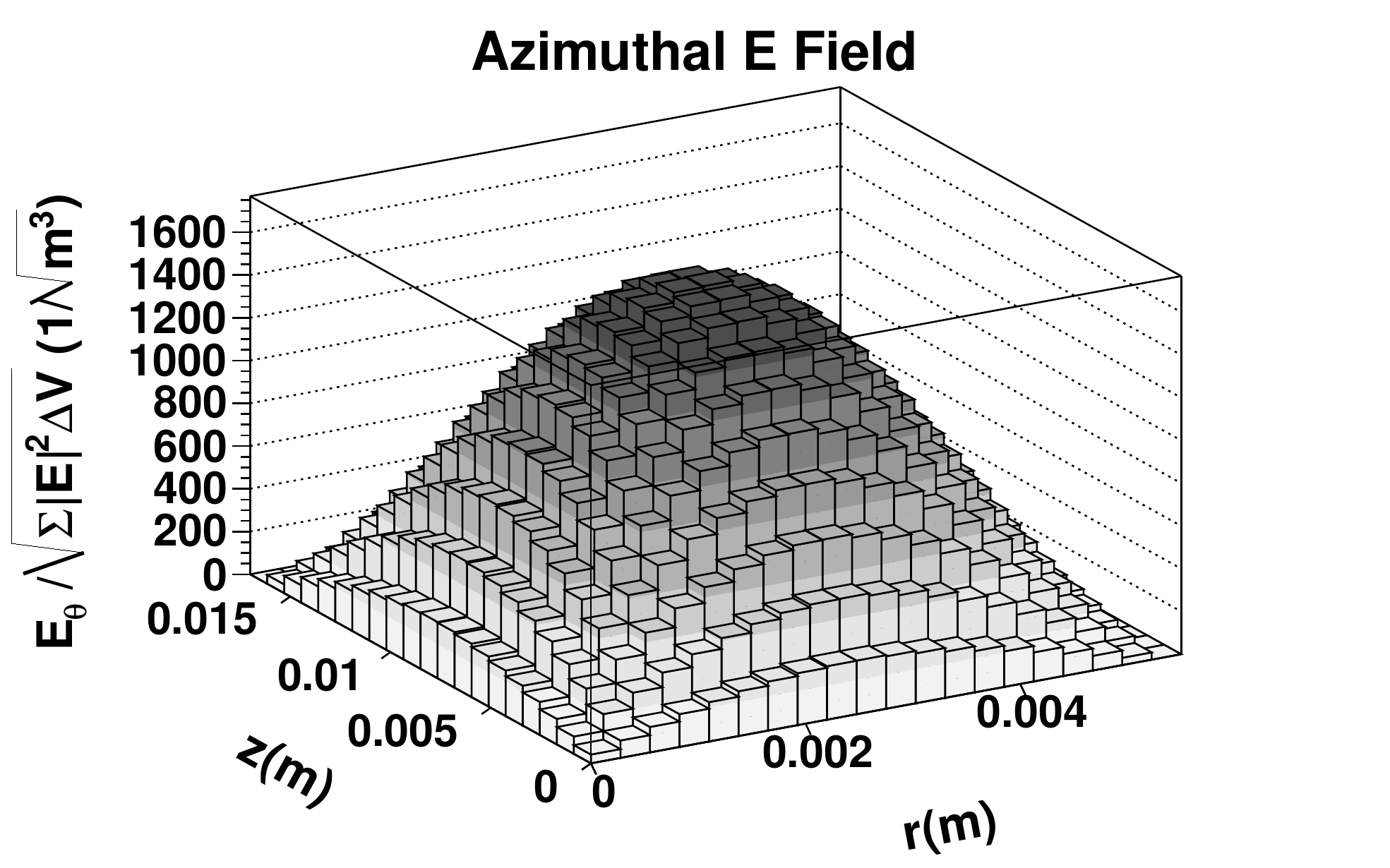}
\caption{Two important components of the fields in the TE$_{011}$
cavity.  The left panel shows B$_z$, the critical component in
Equation~\ref{eq:Cprime_lmn}.  The right panel depicts E$_\theta$
which is the field that enters into the numerator of
Equation~\ref{eq:geomfact}. \label{fig:Fields}}
\end{figure}

Inside the signal cavity, two components of the electric and
magnetic fields are important in this experiment.  The first is the
field aligned with the external B--field in the {\bf z} direction,
which in the case of the TE$_{011}$ mode is theoretically $B_z =
k_r/k J_0(k_r r) sin(k_z z)$ while $E_z=0$. In practice $B_z$ is
perturbed by its coupling to the waveguide. Its magnitude in the
cavity is simulated in accordance with the actual geometry using the
HFSS simulation software in ANSYS$^\circledR$ 
Academic Research Release 12.1.2 and is plotted in the left
panel of Figure~\ref{fig:Fields}. In the second signal
cavity~\cite{malagon} supporting the TM$_{020}$ mode, $B_z=0$; the
diameter is 15~mm and the height is 9~mm.  The axially--aligned
electric field takes the analytic form $E_z = k_r/k J_0(k_r r)$ and
is not plotted.

The second important field component to consider is the azimuthal
electric field, or $E_\theta = J_0^\prime(k_r r) sin(k_z z)$.  It is
the only non--zero component of the electric field in the
cylindrical TE$_{011}$ mode.  The right panel of
Figure~\ref{fig:Fields} shows its analytical form as a function of
height and radius in an ideal cavity.

\subsubsection{Drive Cavity}
The drive cavity runs in the TE$_{011}$ mode at the same resonant
frequency as the TE$_{011}$ signal cavity, at room temperature. Like
the signal cavity its inner volume is on the order of 1~cm$^3$, with
diameter 12~mm and height 17~mm. It is typically driven by a narrow
monoenergetic signal with $\lesssim$1~Watt average power which
allows for thermal equilibrium without cooling. While its resonant
frequency can in principle be tuned by temperature adjustment using
water circulation, it is generally more practical to wait for the
drive cavity to reach thermal equilibrium and then tune the signal
cavity such that their frequencies match.

\subsubsection{Field Overlap Integrals}\label{sec:overlap}

The sensitivity of the experiment to new physics is governed partly
by the orientation of the electric and magnetic fields inside the
resonant cavity.  Each of the two running modes, and each of the two
signal cavities, has a field overlap integral that defines its
particular ability to detect signals driven by interactions with new
light particles.

\begin{figure}[h]
\hspace{1.25in}\includegraphics[width=3.0in]{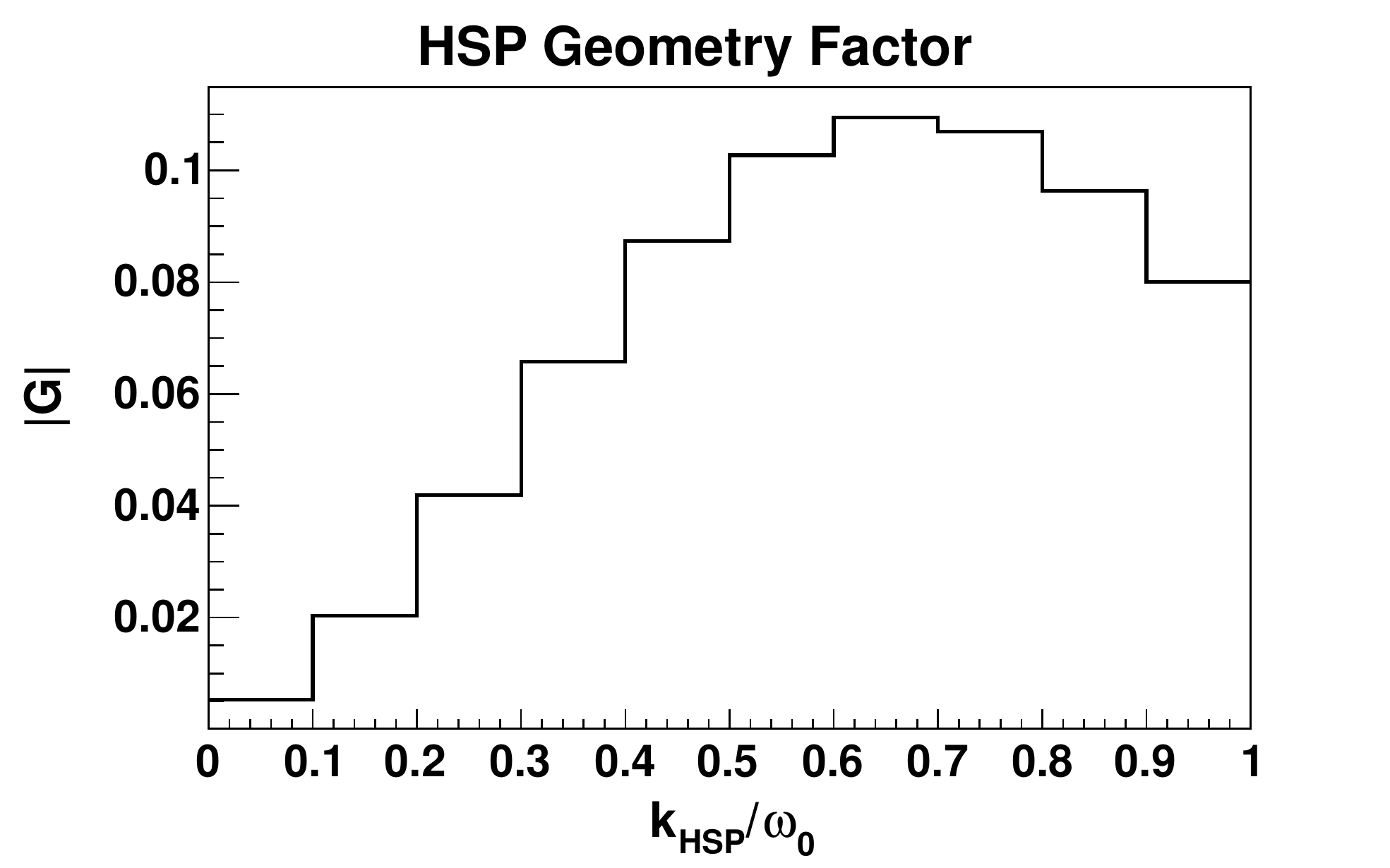}
\caption{Plot of the field overlap integral for hidden sector
photons (HSP), $|G|$~\cite{Jaeckel:2007ch}, against the wavenumber
k$_{HSP}$.\label{fig:geomfact}}
\end{figure}

The running mode with two TE$_{011}$ cavities side by side is
optimized to find oscillations between hidden sector photons and
Standard Model photons.  The sensitivity of the result depends on
the overlap integral in~\cite{Jaeckel:2007ch}
\begin{equation}
G\equiv \omega_o^2\int_{V^\prime}\int_{V}d^3{\bf x}d^3{\bf y}
\frac{\text{exp}(ik|{\bf x}-{\bf y}|)A({\bf y})A^\prime({\bf x})}{4\pi|{\bf
x} - {\bf y}|} \label{eq:geomfact}
\end{equation}
where $\omega_0$ is the drive frequency and $k$ is the wavenumber of
the hidden sector photon.  Following the steps
in~\cite{Jaeckel:2007ch} $E=-dA/dt$.  Taking the spatial part of
$E_\theta({\bf x},t)$ in the TE$_{011}$ cavities from the right
panel of Figure~\ref{fig:Fields}, $|G|$ is computed numerically for
the two side-by-side cavities in this experiment.
Figure~\ref{fig:geomfact} shows the result.

Alternatively, in the case of the listening mode that utilizes just
one signal cavity, the overlap integral $C_{lmn}$ is defined with
respect to the direction of the strong external magnetic field ${\bf
\hat B}$. It is first described
in~\cite{Sikivie:1983ip,Sikivie:1985yu} to quantify the response of
a resonant cavity to a pseudoscalar axion in a strong magnetic
field:
\begin{equation}
C_{lmn}\equiv \frac{\left|\int_V d^3x {\bf E}\cdot{\bf \hat B}
\right|^2} {V\int_V d^3x \epsilon_r |{\bf E}|^2}, \label{eq:overlap}
\end{equation}
where {\bf E} is the electric field in the cavity and $V$ is the
volume.  In the case of the TM$_{020}$ cavity employed in this
experiment, $C_{lmn}$ is on the order of 0.1~\cite{malagon}.

For the signal cavity that runs in the TE$_{011}$ mode, $C_{lmn}$ is
identically zero because $E_z$=0.  However there is a small overlap
$C^\prime_{lmn}$ in the case of scalar axion--like particles,
adapting the calculation from Eq.~\ref{eq:overlap} as
\begin{equation}
\label{eq:Cprime_lmn}
 C^\prime_{lmn}\equiv \frac{\left|\int_V d^3x
{\bf B}\cdot{\bf \hat B} \right|^2} {V\int_V d^3x \frac{1}{\mu_r}
|{\bf B}|^2}.
\end{equation}
The overlap is small at $O$(10$^{-6}$) because of the behavior of
the Bessel function $J_0(k_r r)$ in the axial magnetic field,
plotted on the left in Figure~\ref{fig:Fields}.  The field is
positive near the center of the cavity, and negative toward the
outer rim.  When integrated over the cylindrical volume, ${\bf
B}\cdot{\bf \hat B}$ decreases.

\subsection{Receiver}

\subsubsection{Cryogenic amplifier}
The first and most critical component in the receiver chain is a
low--noise, broadband, high--electron--mobility transistor (HEMT)
amplifier~\cite{weinreb}.  It typically operates at frequencies of
11--26~GHz but is routinely tested for use up to 40~GHz. In this
experiment the HEMT is cooled to 7~K and measures broadband noise
near 34 GHz.  Its specified noise temperature measured at 22~K is
approximately 35~K. In Section~\ref{sec:noisecalibration} the noise
temperature of the HEMT is found to be near 20~K when it is cooled
to 7~K.  While there are actually two HEMT amplifiers connected in
series inside the cryostat for adequate gain, it is the first HEMT
that dictates the noise temperature of the system.  The latter
remains true as long as its output power is large compared to the
noise temperature of the next amplifier.

\subsubsection{Room--Temperature electronics}
The room--temperature receiver uses a triple heterodyne technique to
mix the RF signal down from 34 GHz to baseband.  The block diagram
is included in Figure~\ref{fig:circuit}.  The three--stage design
has been chosen to avoid possible problems with crosstalk related to
high amplification at 34~GHz.  Broadband noise is limited at each
stage by bandpass filters placed before and after the amplifiers,
thereby avoiding saturation.

Power at the image frequencies is suppressed by more than 100 dB
with the bandpass filters (BPFs) that sit before the first two
mixers.  For example, if RF frequency f$_1$ mixes with LO frequency
f$_2$, then the intermediate frequency (IF) is f$_1$-f$_2$ and the
image lies at 2f$_2$-f$_1$. It is this image frequency that must lie
outside the passband of the BPF that precedes each mixer. If it were
not suppressed then the noise power would increase by a factor of 2
after the mixer. In the case of the third mixer, a different
technique is used to suppress the image power.  The outputs of the
mixer in the baseband are separated into the in--phase ($I$) and
quadrature ($Q$) voltages. The image power in the baseband is
suppressed by selecting the sign of the complex phase
$\phi=tan^{-1}(Q/I)$.

The first six harmonics of each LO are also excluded from the
passband of the receiver. For example, the second LO in the chain
oscillates at 3.5 GHz, so its first 6 harmonics are 7.0~GHz,
10.5~GHz, 14.0~GHz, 17.5 GHz, 21.0 GHz, and 24.5 GHz.  The frequency
plan is chosen so that these harmonics do not propagate through the
electronics chain into the baseband. If a harmonic did pass into the
baseband, it could mimic a real signal and cause difficulty in the
offline analysis.  In spite of the above it should be noted that
harmonics with even higher orders are still expected to be present
in the data; care is required, as always, in discriminating between
a real signal and a systematic or environmental feature.

As a preliminary check of the receiver's performance, the output of
the room temperature chain has been checked against its
specifications at the first two stages with a 50~$\Omega$
termination at the input.  The specified noise figure and gain of
the components in the room--temperature chain are cascaded according
to the Friis formula for total noise factor
\begin{equation}
F=F_1 + \frac{F_2-1}{G_1} + \frac{F_3-1}{G_1G_2} + \cdot\cdot\cdot,
\end{equation}
where $F$ represents the total noise factor, $F_i$ is the specified
noise factor for component $i$, and $G_i$ is the gain of component
$i$. The expected noise power in the chain at any given point $i$ is
$N_S+G+10log_{10}(F)$~dBm/Hz where $N_S$ is the noise power of the
source at the input to the chain and $G$ is the total gain. The
predicted noise power is then compared with the measured noise power
in Table~\ref{Tab:receiver}, with reasonable agreement.
\begin{table}
\centerline{\begin{tabular}{|l|r|r|} \hline
component &  measured      &  predicted  \\
          & (dBm/Hz)       & (dBm/Hz) \\
\hline
300K 50$\Omega$ term.   & $<$-150   & -174 \\
mixer 1  &     $<$-150  &  -173     \\
IF1 amp      &        -112     &  -112     \\
mixer 2   &       -116    &  -118 \\
IF2  amp      &    -85     &  -85  \\
\hline
\end{tabular}}
\caption{Table of noise power measured after the first two stages of
the microwave receiver, compared with values expected from the Friis
formula for cascaded noise power.} \label{Tab:receiver}
\end{table}

\begin{figure}
\hspace{-0.2in}
\includegraphics[width=3.0in]{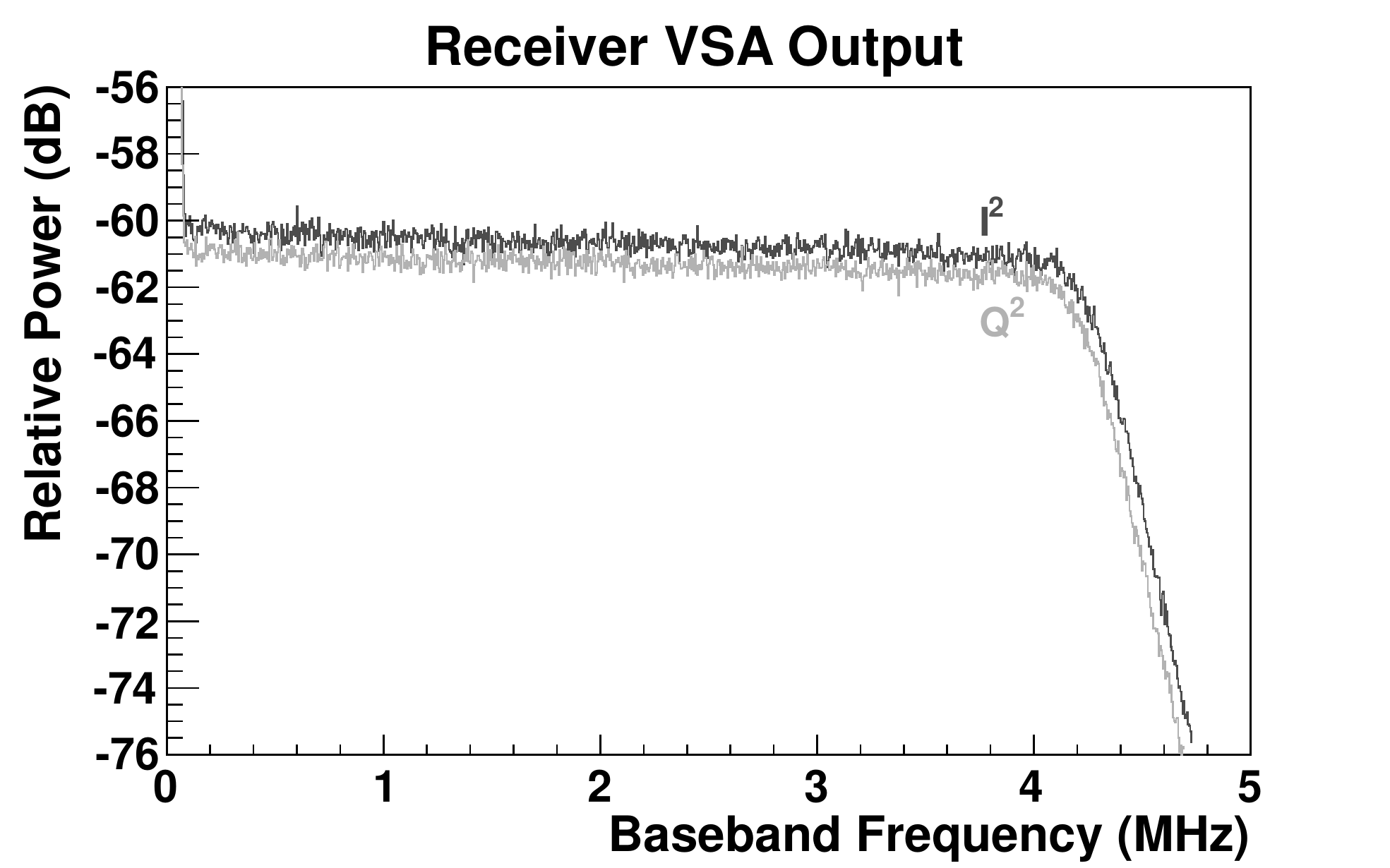}
\includegraphics[width=3.0in]{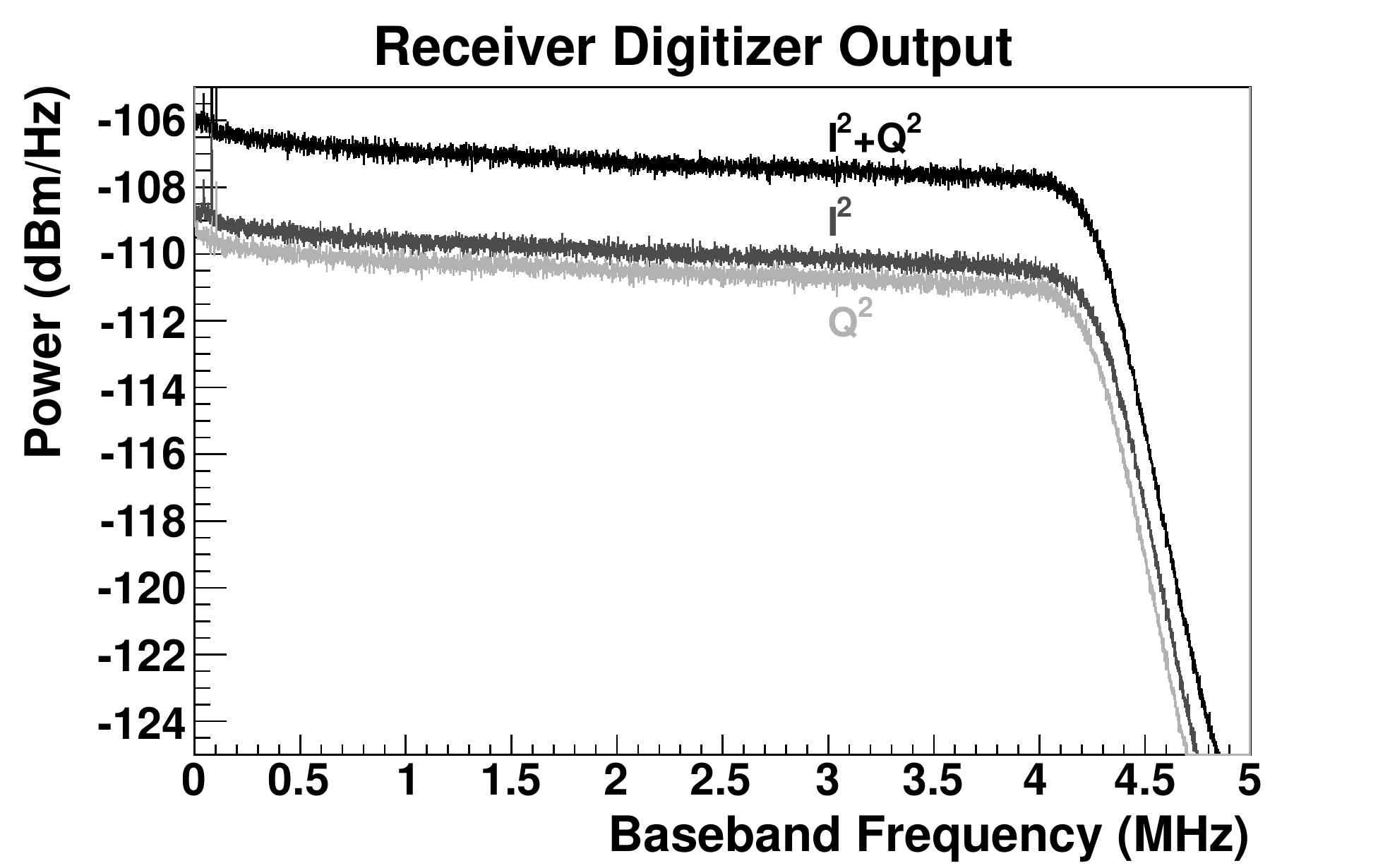}
\caption{Plots comparing the output of the receiver chain in the
baseband.  The left panel shows the outputs in $I$ and $Q$ measured
with the spectrum analyzer.  The right panel shows the outputs after
the digitizer and offline analysis.\label{fig:VSAOutput}}
\end{figure}

As shown in Figure~\ref{fig:circuit}, a low--pass filter typically
precedes the digitizer in each of the two paths $I$ and $Q$.
Figure~\ref{fig:VSAOutput} shows the baseband power in $I$ and $Q$
measured with an Agilent PXA N9030 spectrum analyzer (left panel)
and with the digitizer (right panel). The absolute magnitudes
derived from each instrument are not expected to be identical due to
losses that are characteristic of each digitizer.  However the
relative magnitudes of the power in I and Q are expected to match.

\section{Measurements}

\subsection{Noise Calibration}
\label{sec:noisecalibration}
 The system noise temperature $T_{sys}$
is the figure of merit that drives the statistical uncertainty
$\sigma_T$ in the measurements, according to the Dicke radiometer
equation~\cite{Dicke}
\begin{equation}
\sigma_T = \frac{T_{sys}}{\sqrt{\Delta \nu \tau}} \label{eq:dicke}
\end{equation}
where $\Delta\nu$ is the resolution bandwidth and $\tau$ is the
integration time.  From the equation, $\sigma_T$ decreases with
longer integration times and increases with narrower bandwidths. In
this experiment, $T_{sys} = T_{\text{HEMT}} + T_{th}$, where $T_{\text{HEMT}}$ is
the noise temperature of the HEMT and $T_{th}$ is the physical
temperature of the HEMT's chassis and connectors.

The total uncertainty $\sigma_{tot}$ in the measurements is driven
by both the statistical uncertainty $\sigma_T$ and the systematic
uncertainties $\sigma_{sys}$.  Effectively the two sources of error
are combined in quadrature as
\begin{equation}\label{eq:totalsigma}
\sigma_{tot} = \sqrt{\sigma_T^2 + \sigma_{sys}^2}.
\end{equation}
From Equation~\ref{eq:totalsigma}, $\sigma_{tot}$ improves with
integration time only until $\sigma_T<\sigma_{sys}$.  This means
that in a system with high $T_{sys}$ and small $\Delta\nu$, long
integration times ($\sim$hours) can be beneficial.  Conversely, for
low $T_{sys}$ and large $\Delta\nu$, the condition
$\sigma_T<\sigma_{sys}$ happens quickly and $\sigma_{tot}$ may be
optimal after relatively short integration times ($\sim$seconds).

\subsubsection{Noise power density and twice power methods}
\label{sec:twicepower} 
\begin{figure}[h]
\hspace{1.25in}\includegraphics[width=3.0in]{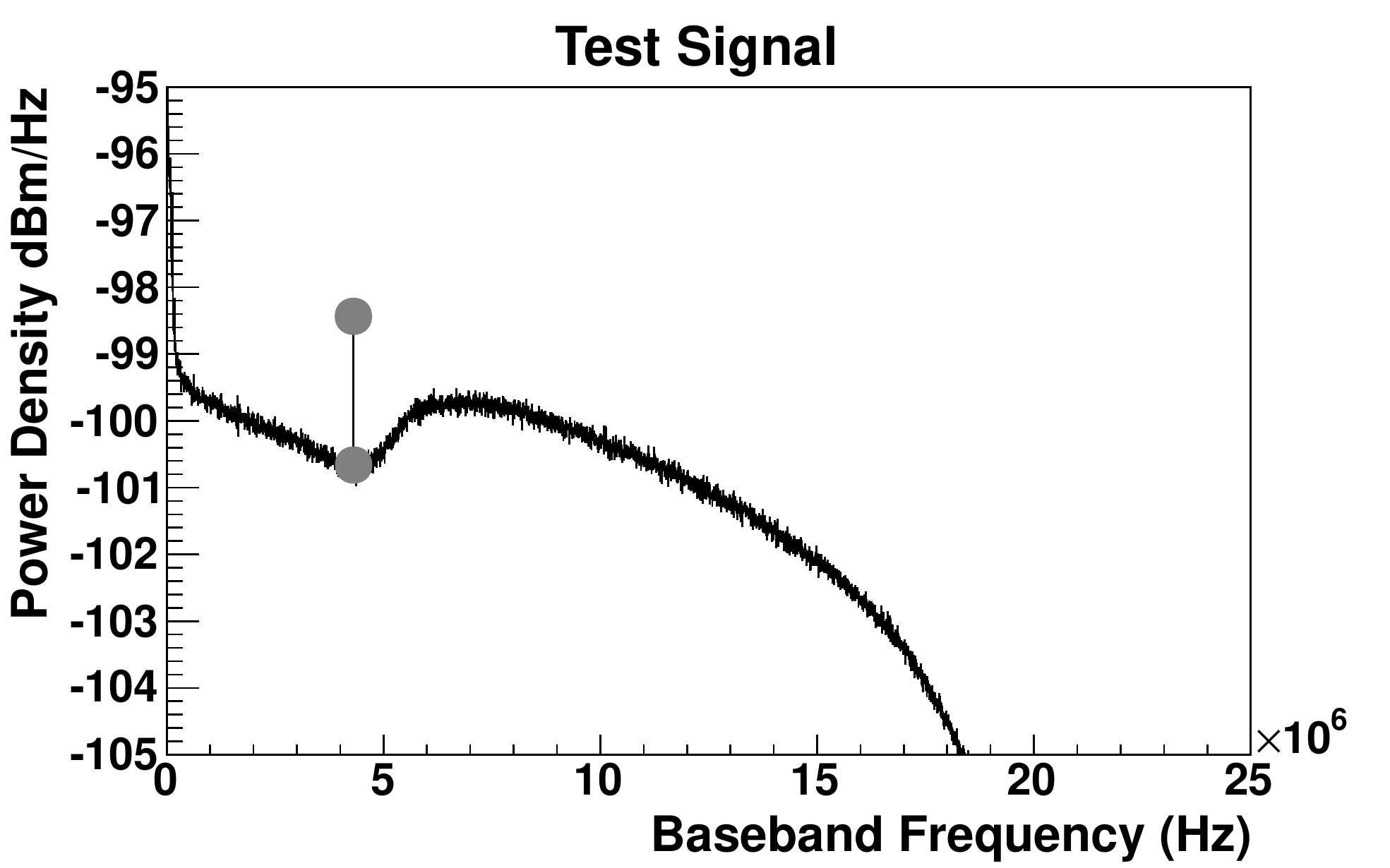}
\caption{Broadband noise measured in the baseband without the low
pass filters in place.  The spike at 4~MHz is a test signal that was
sent into the cavity through the weakly coupled calibration
port.\label{fig:testsignal}}
\end{figure}

The system noise temperature is measured
using several approaches, and the results are compared.  First is a
measurement of the total output noise power with a matched
50~$\Omega$ terminator at the input to the cold HEMT.  In this
approach, the mean output power divided by the gain of the
electronics chain is defined as $T_{sys}$.  For this purpose the
gain of the electronics chain is measured with a test signal that is
sent through the electronics chain by way of the calibration
waveguide. However, a problem with this technique is that the power
contained in the test signal is reduced by the line loss, or the
loss in the waveguide and through both ports in the cavity at
cryogenic temperatures. Because these behaviors are hard to
characterize in the presence of reflections and standing waves, the
gain inferred from a test signal is reported only as a function of
the line loss. Figure~\ref{fig:testsignal} shows an example of a
test signal sitting on a background of amplified thermal and
electronic noise.

\begin{figure}[h]
\hspace{-0.2in}
\includegraphics[width=3.0in]{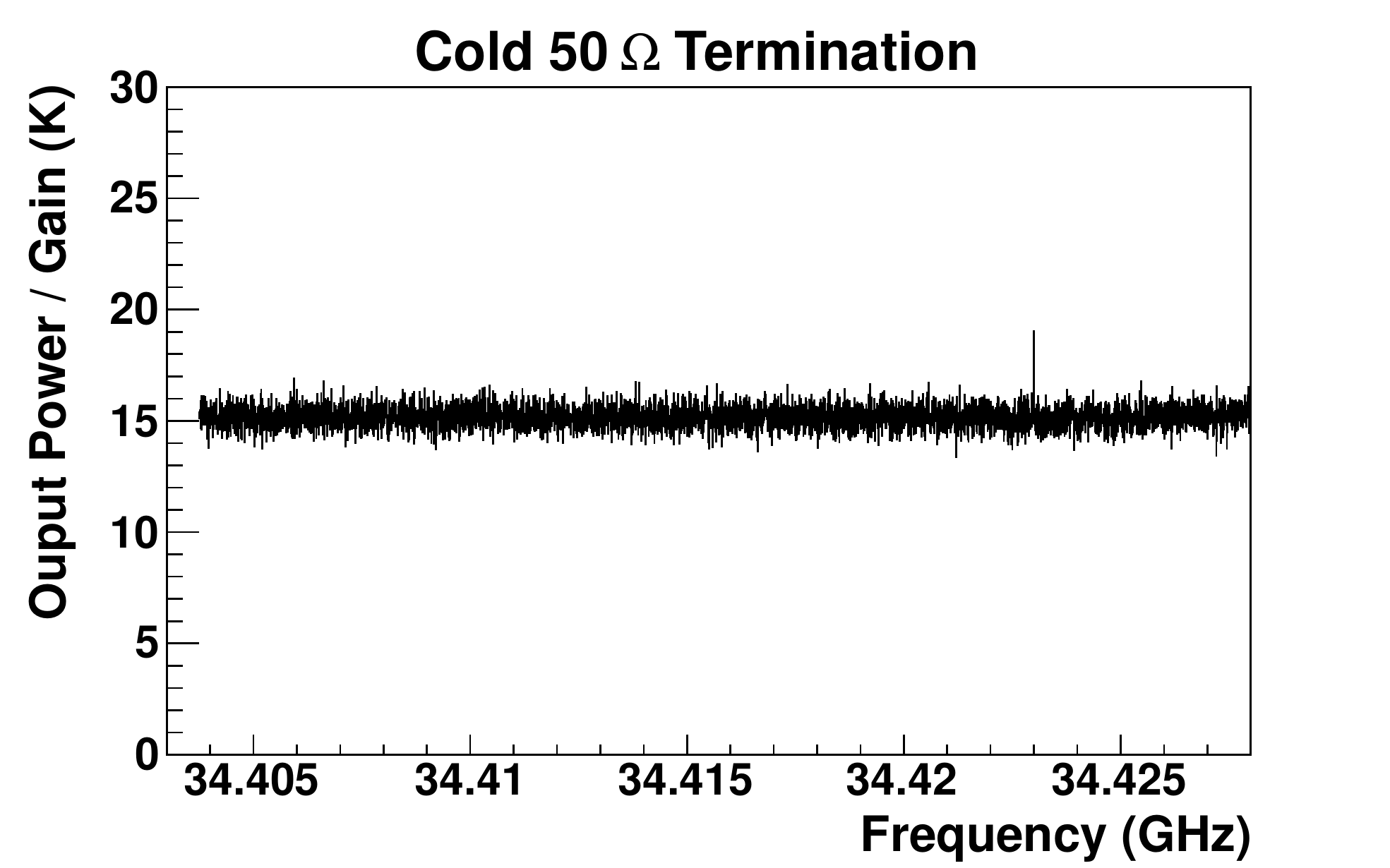}
\includegraphics[width=3.0in]{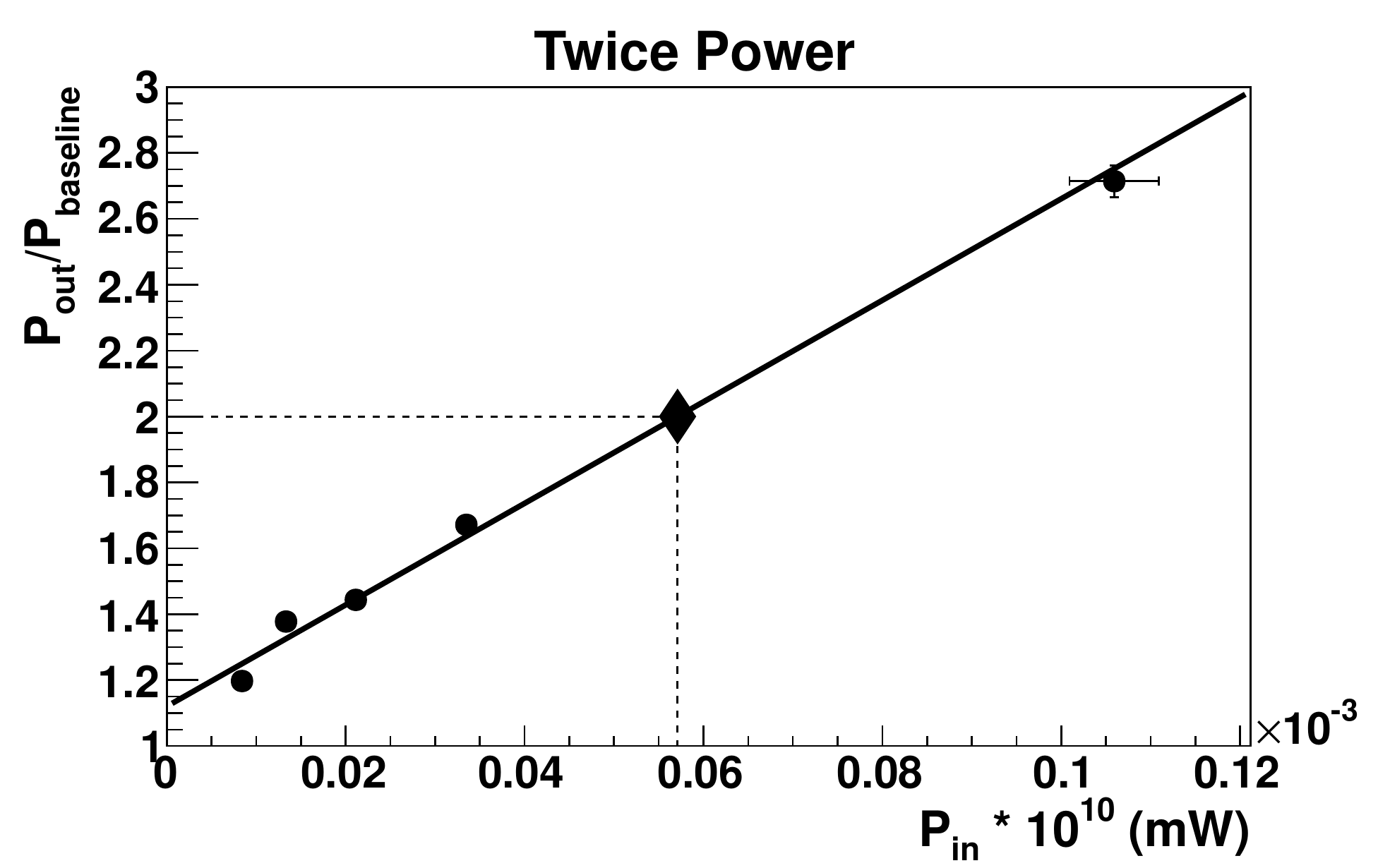}
\caption{The left panel shows broadband noise from the receiver,
corrected for the hardware transfer function, with a 7~K matched
50~$\Omega$ termination at the input.  The spike at 34.423~GHz is
related to the 10~MHz reference oscillator.  The circular points in
the right panel show the magnitudes of a series of test signals
relative to the surrounding noise, plotted against the presumed
input power $P_{in}$. The input power for which the output power is
doubled is inferred by interpolation, shown as the diamond marker on
the plot.
 The magnitudes of $P_{in}$ have been temporarily scaled by a factor of 10$^{10}$
 to accommodate the precision of the fitting algorithm on the 32-bit computer.\label{fig:noisetemperaturemethods}}
\end{figure}

With the gain defined as above, $T_{sys}$ can be derived as a
function of the line loss.  With a 50~$\Omega$ termination at the
input to the 7~K HEMT, the output power is corrected for the
hardware transfer function and divided by an array of values for the
gain. An example for one value of the gain is shown on the left in
Figure~\ref{fig:noisetemperaturemethods}, and the results for the
other gains are included on the right in Figure~\ref{fig:yfactor}.

The noise factor of the electronics, driven by $T_{\text{HEMT}}$, can also
be deduced with the ``twice--power'' method. In this approach a narrow test
signal with power $P_{in}$ is injected into the electronics until it
sits 2$\times$ higher than than the baseline noise at the output of
the electronics. Then, the noise factor $F \equiv P_{in}/k_BT\Delta\nu$
where $k_BT\Delta\nu$ is the baseline noise at the input.  Both $P_{in}$
and $k_BT\Delta\nu$ are extracted using the gain of the electronics chain,
which as stated above, is still a theoretical function of the line
loss. On the right in Figure~\ref{fig:noisetemperaturemethods} is a
plot of $P_{out}$ against $P_{in}$ for a set of test signals with a
range of magnitudes.  The condition $P_{in}/P_{baseline}\equiv$2 is
determined by interpolation in order to calculate $F$. The noise
factor and noise temperature $T$ of the HEMT are related as
$T=(F-1)T_0$ where $T_0$ is the temperature of the HEMT's chassis.
$T_{sys}$ is then derived by adding the chassis temperature $T_{th}$
to $T$.  The right panel of Figure~\ref{fig:yfactor} includes
$T_{sys}$ derived from $F$ as a function of the line loss.  As shown
on the plot, the results agree well with $T_{sys}$ taken from the
measurements of the mean broadband noise from the matched
50~$\Omega$ termination.

\subsubsection{Y-Factor Method}
The uncertainty caused by the line loss described in the previous
section is removed with a Y-factor measurement.  In this approach,
the frequency--dependent noise temperature of the amplifier is
derived from the ratio of two spectra with different input
temperatures while the temperature of the HEMT is held constant. The
first spectrum is taken using a cold source ($\sim$7~K) as the
input, and the second with a warm source ($\sim$28~K).  The
temperature of the HEMT is reasonably constant, ranging between 4.5
and 8~K.  The temperatures are determined with the thin--film
resistance cryogenic temperature sensors.  The noise temperature of
the amplifier is defined as the X--intercept on a graph of output
power plotted against source temperature.  The data are collected
using an Agilent 9030A spectrum analyzer.

During the measurement, a Cu block is in thermal contact with a
50~$\Omega$ termination that is connected to the input of the HEMT
using a 5~cm long 0.085$^{\prime\prime}$ diameter cable with stainless steel
jacket, PTFE dielectric interior, and stainless steel inner
conductor.  The cable provides some thermal isolation between the
HEMT and the termination.  Additional isolation is achieved by
wrapping the block loosely with mylar lined on the inside with a
layer of Dacron mesh.  For temperature stability, the HEMT is
thermally coupled to a pool of liquid He at the bottom of the
cryostat with an OFHC Cu cold finger.  The temperature of the source
is increased by applying current to a resistor that is in thermal
contact with the Cu block.

\begin{figure}[h]
\hspace{-0.2in}
\includegraphics[width=3.0in]{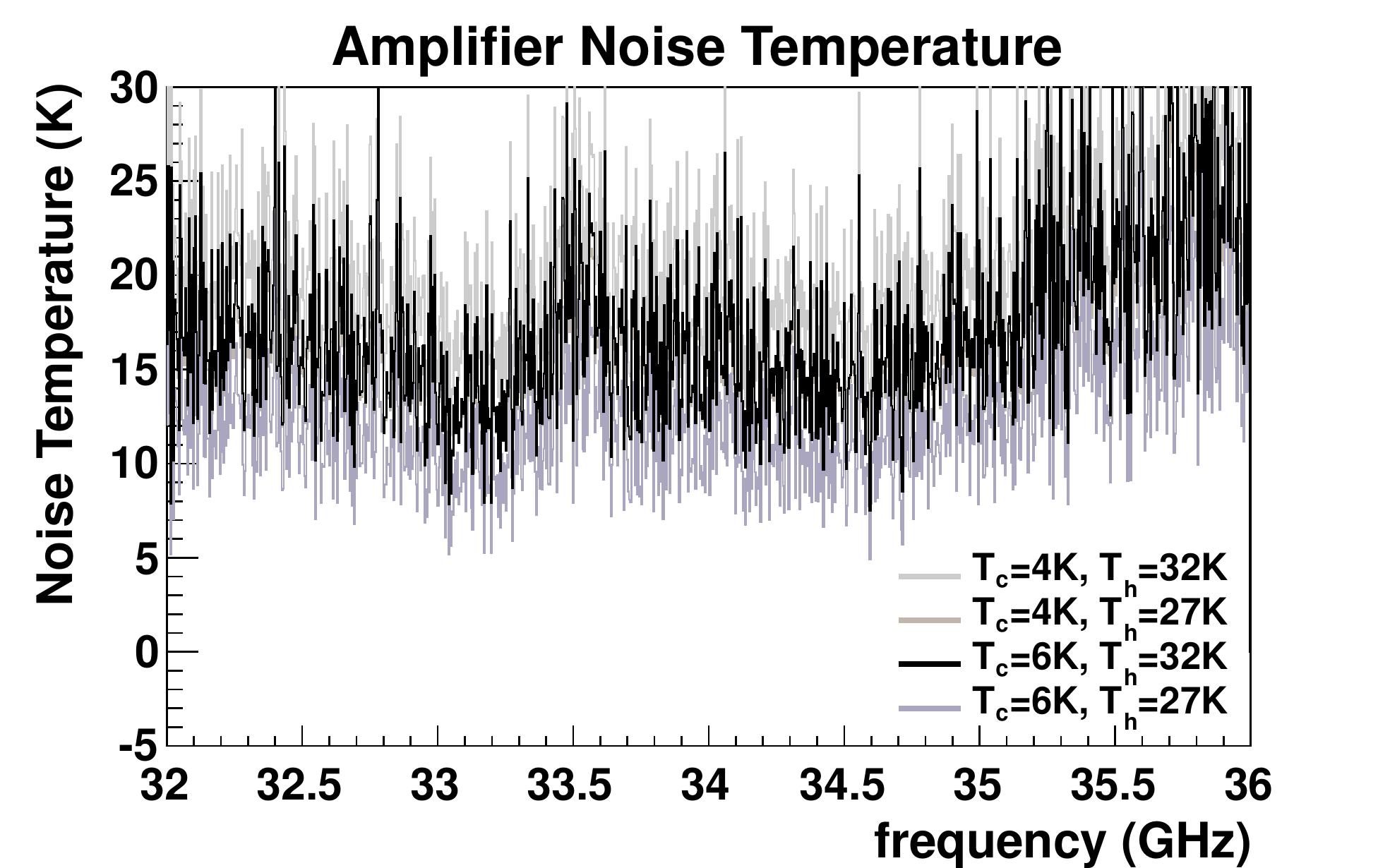}
\includegraphics[width=3.0in]{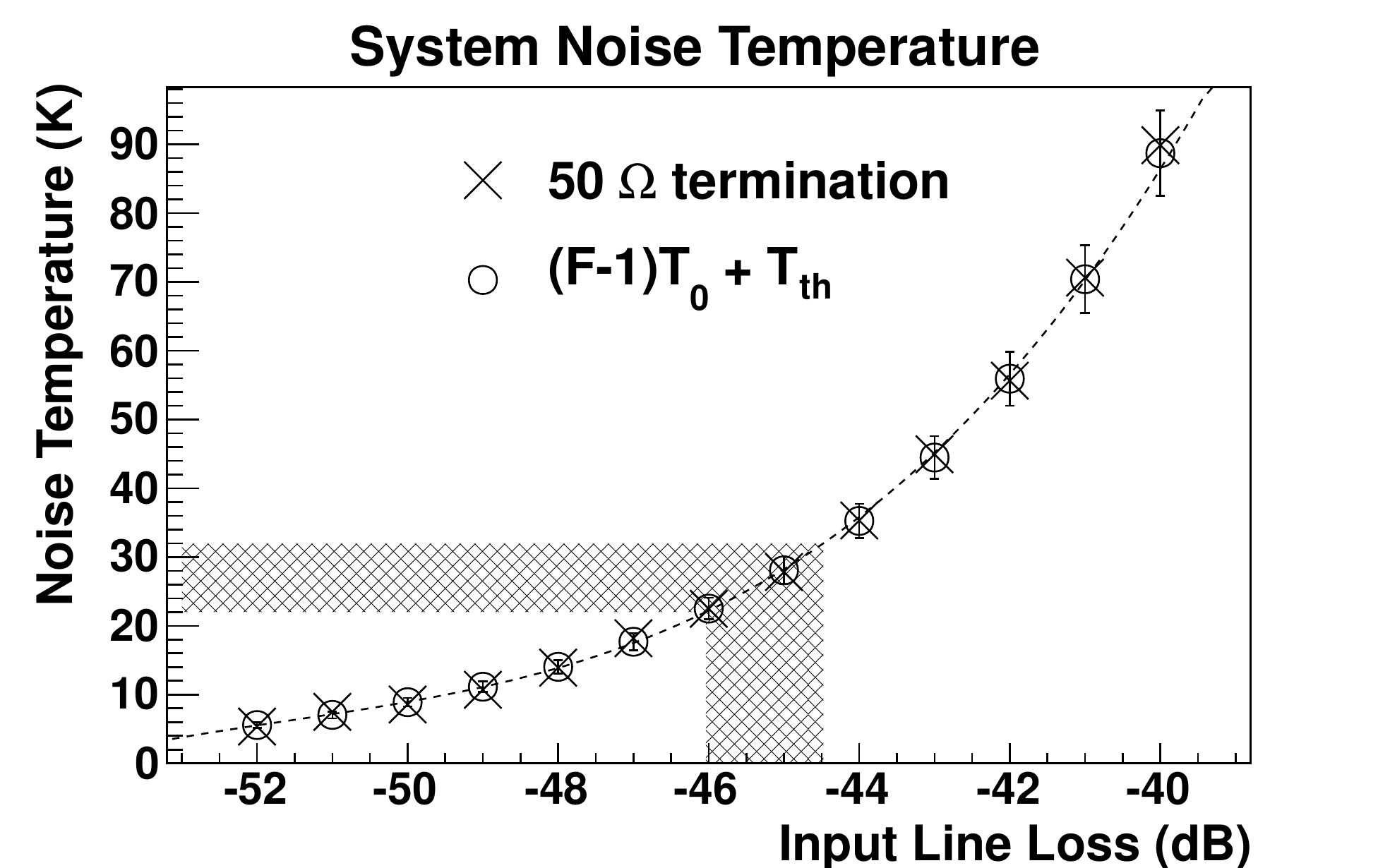}
\caption{The left panel shows the amplifier noise temperature
derived from the Y--factor measurement.  The legend contains several
sets of plausible cold and warm source temperatures T$_c$ and T$_h$.  In the right panel, the
shaded region shows $T_{sys}$ inferred from the left panel assuming a
cryostat temperature $T_{th}$ of 7~K. The right panel
also depicts $T_{sys}$ as calculated from measurements with a
50~$\Omega$ termination and from the twice-power
method.\label{fig:yfactor}}
\end{figure}

Figure~\ref{fig:yfactor} summarizes the results of the Y-factor
measurement.  The left panel contains the extracted noise
temperature of the electronics, driven by the HEMT.  The largest 
sources of uncertainty come from imperfect thermal contact between the
temperature sensors and the warm and cold terminations, and from
unwanted heating of the amplifier during the measurement.  Additional
errors arise from a lack of thermal
equilibrium between the Cu block and the 50~$\Omega$ termination, time
lag between heating of the source and the data collection, and uncertainty
in the amount of RF power lost in the stainless steel cable.
The loss in the stainless steel cable is estimated to be 0.6~dB, adjusted
downward from 0.9~dB at room temperature according to the expected lowering
of electrical resistivity at 5~K~\cite{nist260-46}.
Considering all of the above, $T_{\text{HEMT}}$ is probably near
20$\pm$5~K. The right panel of Figure~\ref{fig:yfactor} shows
$T_{sys}=T_{\text{HEMT}}+T_{th}\cong$27$\pm$5~K taken from the Y-factor
measurement, plotted simultaneously with the measurements discussed
in Section~\ref{sec:twicepower}.  As a result of the Y-factor
measurement the gain of the system is inferred to be
84.0$\pm$1.0~dB.

The measured noise temperature of the amplifier falls within a factor of 12$\pm$3 of the
standard quantum limit at 34~GHz $hf/k_B=$1.6~K.  While this is not atypical for
a HEMT amplifier whose semiconductor structures are built from
InP~\cite{reid2008}, it is perhaps notable that the noise performance does
not appear to be significantly degraded by the strong 7~T magnetic field.  
Daw and Bradley~\cite{Daw:1997} found that the noise performance of 
a HEMT built with GaAs/AlGaAs and operating at 683~MHz was degraded by the presence 
of an ambient 3.6~T magnetic field.  The effect was found to be highly dependent on 
the orientation of the amplifier relative to the field.  A quantitative account for the 
behaviors was given in terms of the electrons' trajectories across the two-dimensional
electron gas between semiconductor layers~\cite{Daw:1997}.

For the case of the InP HEMT in the present experiment, there is no obvious 
reason why the electrons in the two-dimensional gas should not be affected 
by the magnetic field similarly to~\cite{Daw:1997}.  Furthermore, through
engineering constraints the amplifier is oriented relative to the magnetic field such
that if there is such an effect, it should be maximal.  Additional investigation 
is therefore required to fully characterize the noise performance of the InP HEMT 
in the magnetic field, including observations of the current drawn, the gain, and 
the inferred noise temperature.

\begin{figure}
\hspace{1.25in}\includegraphics[width=3.0in]{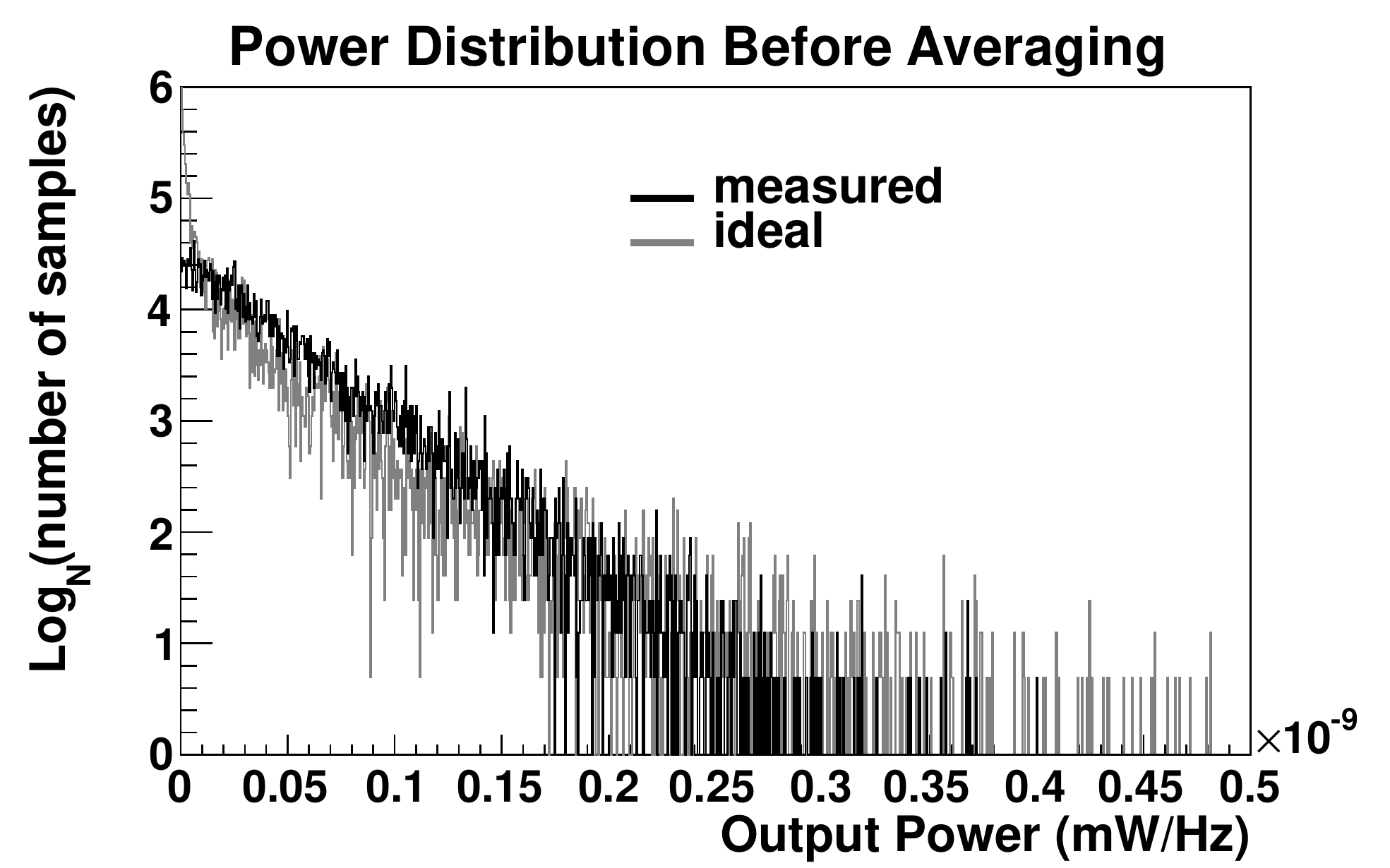}
\caption{Distribution of power measured with the receiver, plotted
simultaneously with the ideal distribution having the same mean.
\label{fig:powerdistribution}}
\end{figure}

As a complement to the above measurement of the mean $T_{sys}$, it is
important to examine the distribution of individual power samples.
Their values are expected to have a predictable behavior when
sampled with an ideal total power radiometer.  The raw sampled
voltages $V$ should follow a Gaussian distribution in number density
centered around 0~V.  The power measurements should fall inside the
same Gaussian, squared~\cite{condon}
\begin{equation}
P(V^2) \sim \frac{1}{V}e^{(-V^2/(2\sigma^2))},
\end{equation}
where $\sigma$ is the standard deviation of the Gaussian and
$\sigma^2$ becomes the mean of the Gaussian squared~\cite{condon}.
Applying the envelope $P(V^2)$ to a random number generator
simulates the output of an ideal radiometer.
Figure~\ref{fig:powerdistribution} shows a plot of the simulated
data compared with real data from this experiment. The means in each
distribution are the same.  From the plot it is apparent that the
qualitative behavior of the two data sets are similar.  Differences
are also present:  The ideal data contain samples at the low and
high energies that are not present in the real data.  These are
attributed to the finite noise floor of the digitizer on the low
end, and to infrequent saturation of the receiver on the high end.
The slight discrepancy near 0.05 mW/Hz is caused by the low energy
tail combined with the arbitrary requirement that the mean be
identical in both data sets.

\subsection{Wave Model}
\begin{figure}
\hspace{1.25in}\includegraphics[width=2.5in]{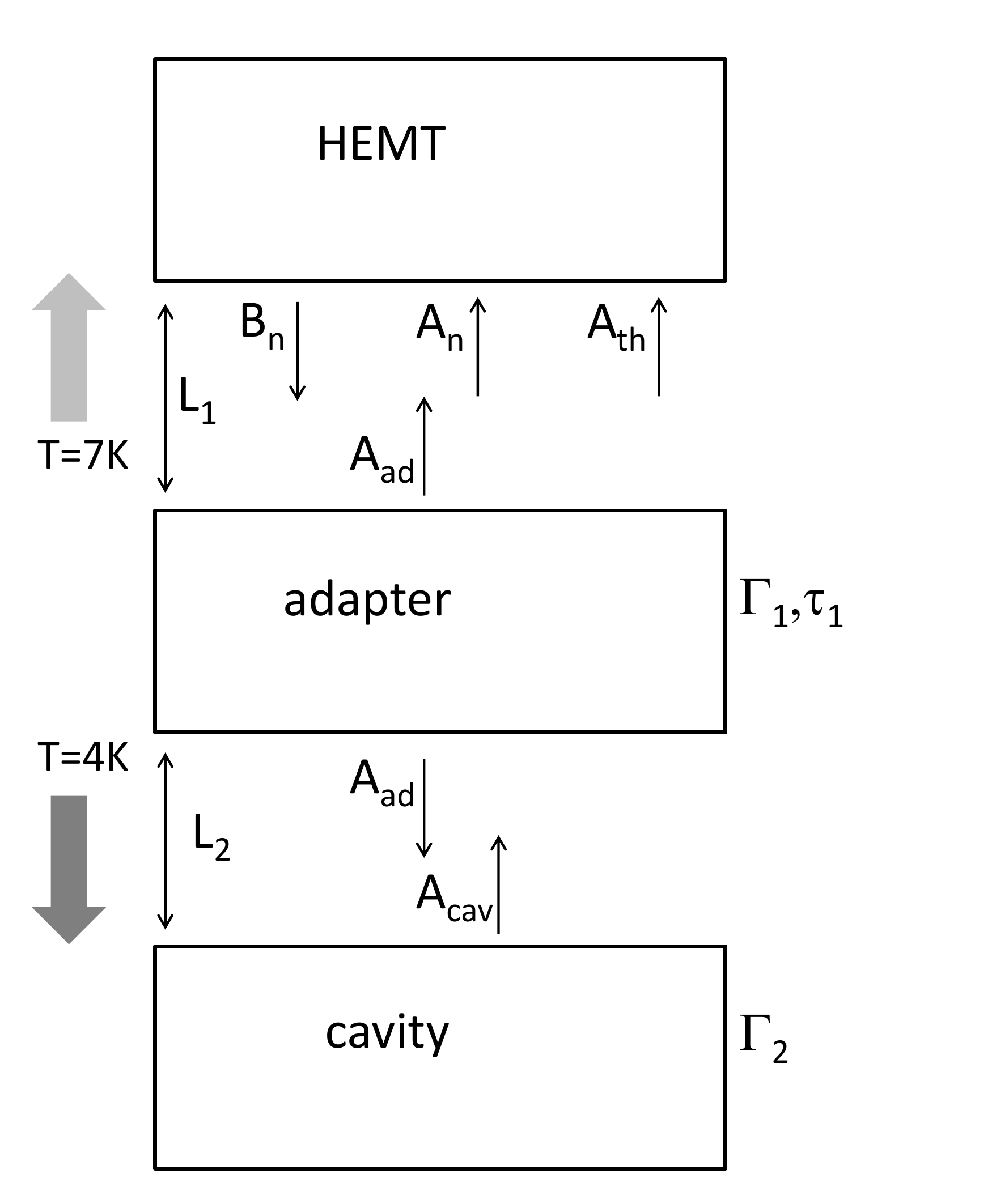}
\caption{Block diagram as the basis for a wave model of the noise
properties of the experiment, after~\cite{Meys}. $L_1$ and $L_2$ are
electrical lengths, and $\Gamma$ and $\tau$ are reflection and
transmission coefficients. The fields $A_n$ and $B_n$ are
characteristic of the amplifier.  All other fields are derived from
physical temperatures.
 \label{fig:wavemodelblock}}
\end{figure}

In the previous section the system noise temperature $T_{sys}$ was
discussed.  In this section $T_{sys}$ will be considered only as it
applies to the system with a resonant cavity at the input. Following
the approach discussed in~\cite{Meys, wedge} for a noisy two--port
device coupled to a noise source, a model of the present experiment
is constructed. Figure~\ref{fig:wavemodelblock} shows a block
diagram of the cavity, WR28--to--coaxial adapter, and the HEMT
amplifier with the noise fields and their directions.  As
in~\cite{Meys} $A_n$ and $B_n$ are the complex ingoing and outgoing
noise waves of the amplifier. $A_{th}$, $A_{ad}$, and $A_{cav}$ are
the waves associated with the physical temperatures of the 3
components, where $|A_{th}|>|A_{ad}|>|A_{cav}|$. The reflection and
transmission coefficients $\Gamma$ and $\tau$ are determined from
the return loss (RL) such that RL(dB) = -20log$_{10}$($|\Gamma|$)
and $|\Gamma|^2+|\tau|^2\equiv$1.  $L_1$ and $L_2$ are electrical
lengths.

Summing the fields at the input to the HEMT gives
\begin{equation}\nonumber
A_n+A_{th}+\Gamma_1B_n + A_{ad} +
\tau_1(A_{cav}+\Gamma_2(B_n\tau_1+A_{ad})),
\end{equation}
which reduces to $A_n+A_{th}+\Gamma_1B_n+\tau_1^2\Gamma_2B_n$
because the HEMT is warmer than the adapter and the cavity.  The
power at the input to the HEMT is taken from the square of the
summed fields.  Following the steps in~\cite{Meys} which are
partially summarized in Equations~\ref{eq:meys}--\ref{eq:meys2},
correlated cross terms retain a phase shift and uncorrelated cross
terms collapse to zero. The noise temperature of the amplifier is
proportional to $|A_n|^2$. The waves $A_n$ and $B_n$ going in and
out of the amplifier are correlated by a phase shift $\phi_c$.   The
electrical lengths $L_1$ and $L_1+L_2$ behave as phase shifts in the
reflection coefficients $\Gamma_1$ and $\Gamma_2$.
\begin{eqnarray}
|A_n|^2 &=& k_B T_a \Delta f\label{eq:meys}\\
|B_n|^2 &=& k_B T_b \Delta f\\
A_n*B_n &=& k_B T_c\Delta f e^{i\phi_c}\\
\Gamma &=& |\Gamma|e^{i\phi_s} \text{ where } \phi_{s1} =
2L_1/\lambda \text{ and } \phi_{s12} =
2(L_1+L_2)/\lambda.\label{eq:meys2}
\end{eqnarray}

\begin{figure}
\hspace{-0.2in}
\begin{minipage}{1.0\textwidth}
\includegraphics[width=3.0in]{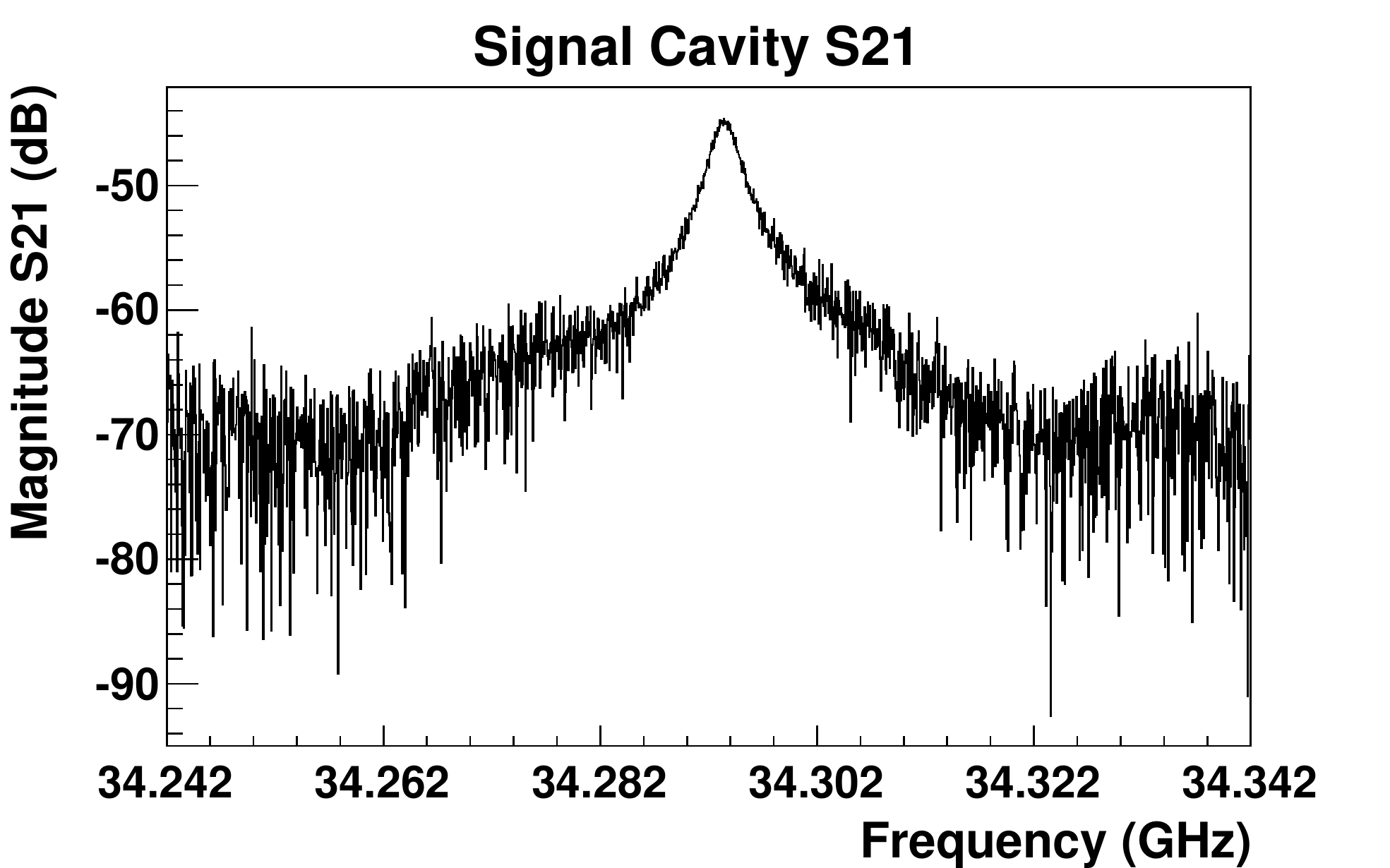}\includegraphics[width=3.0in]{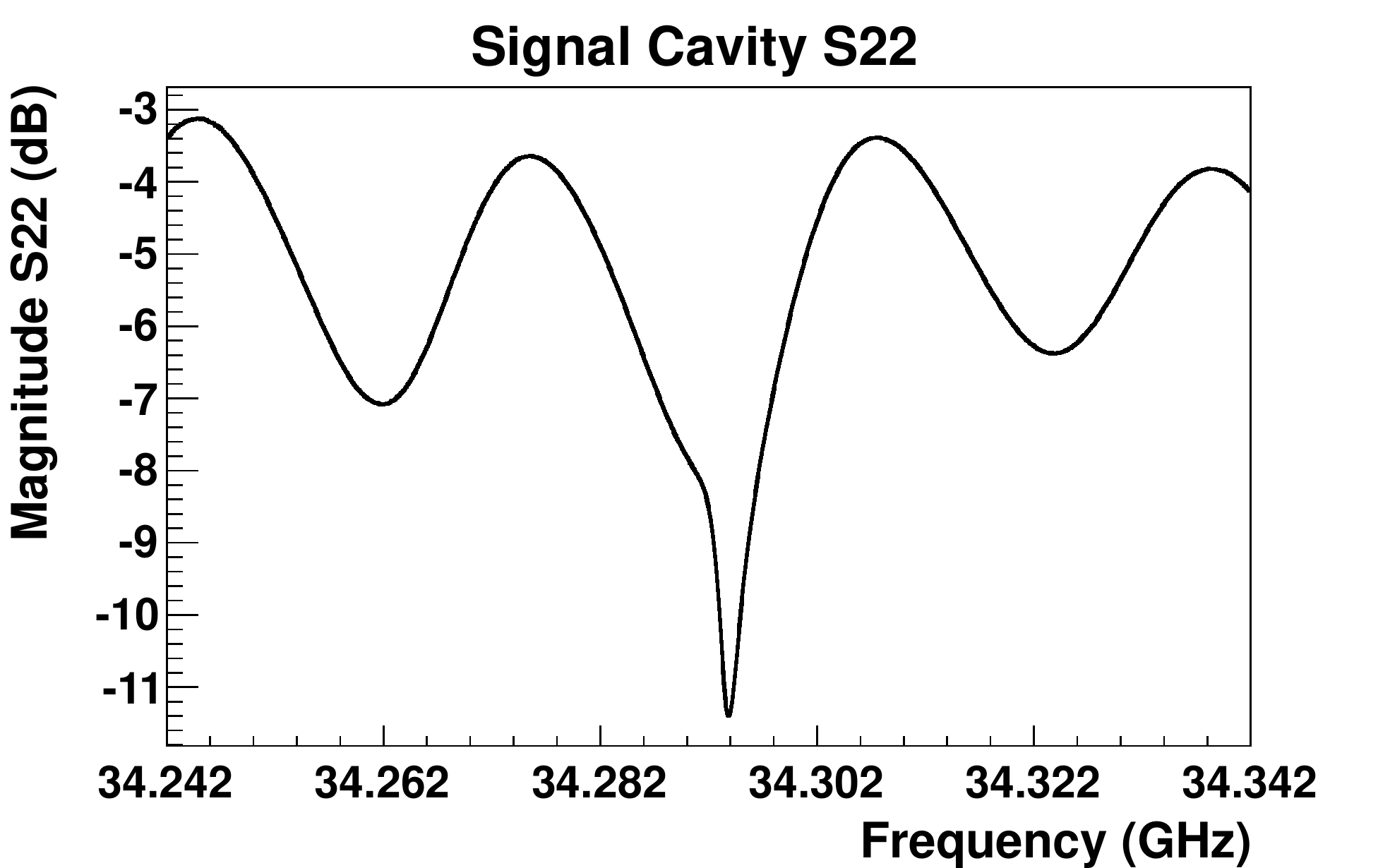}
\includegraphics[width=3.0in]{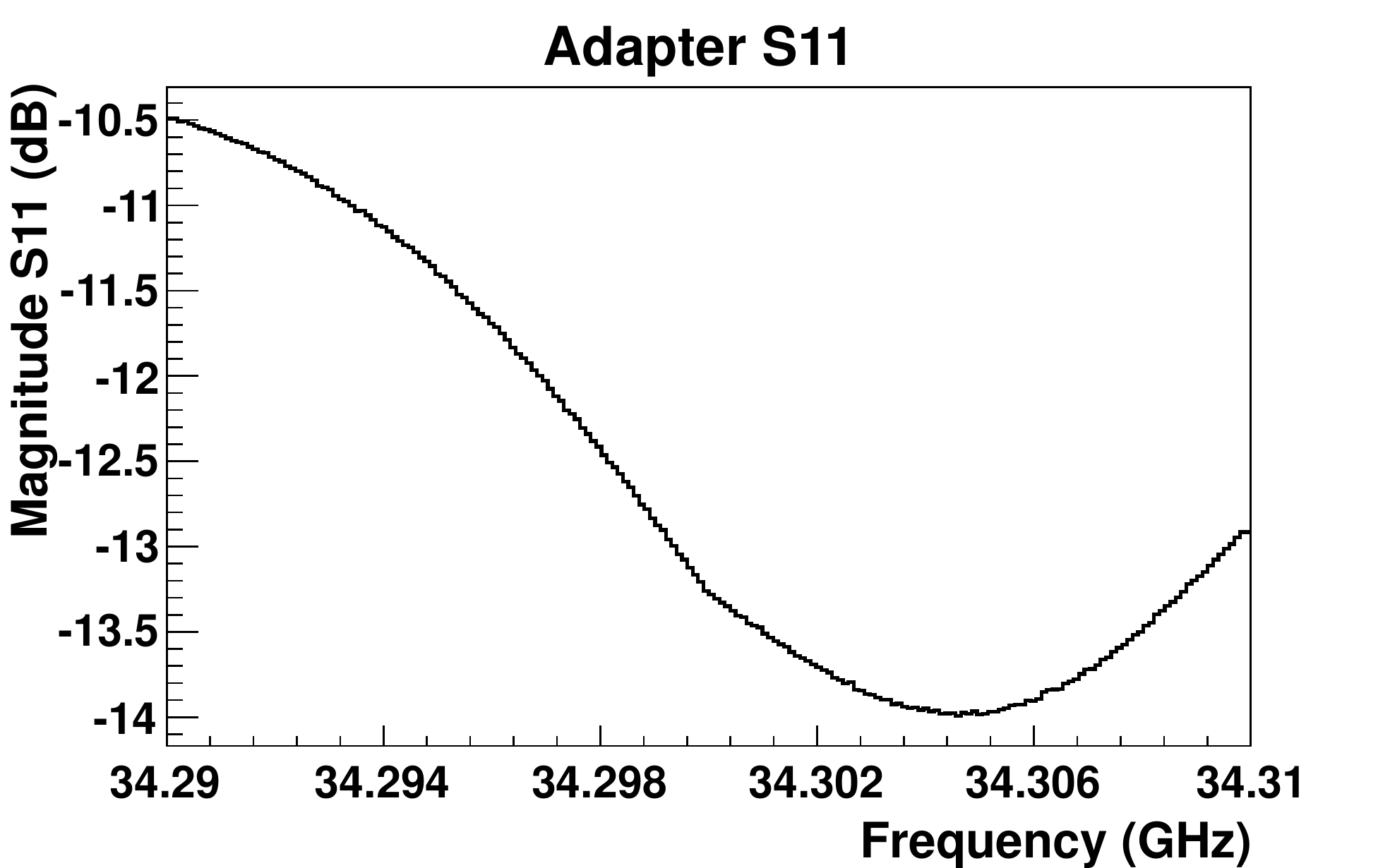}\includegraphics[width=3.0in]{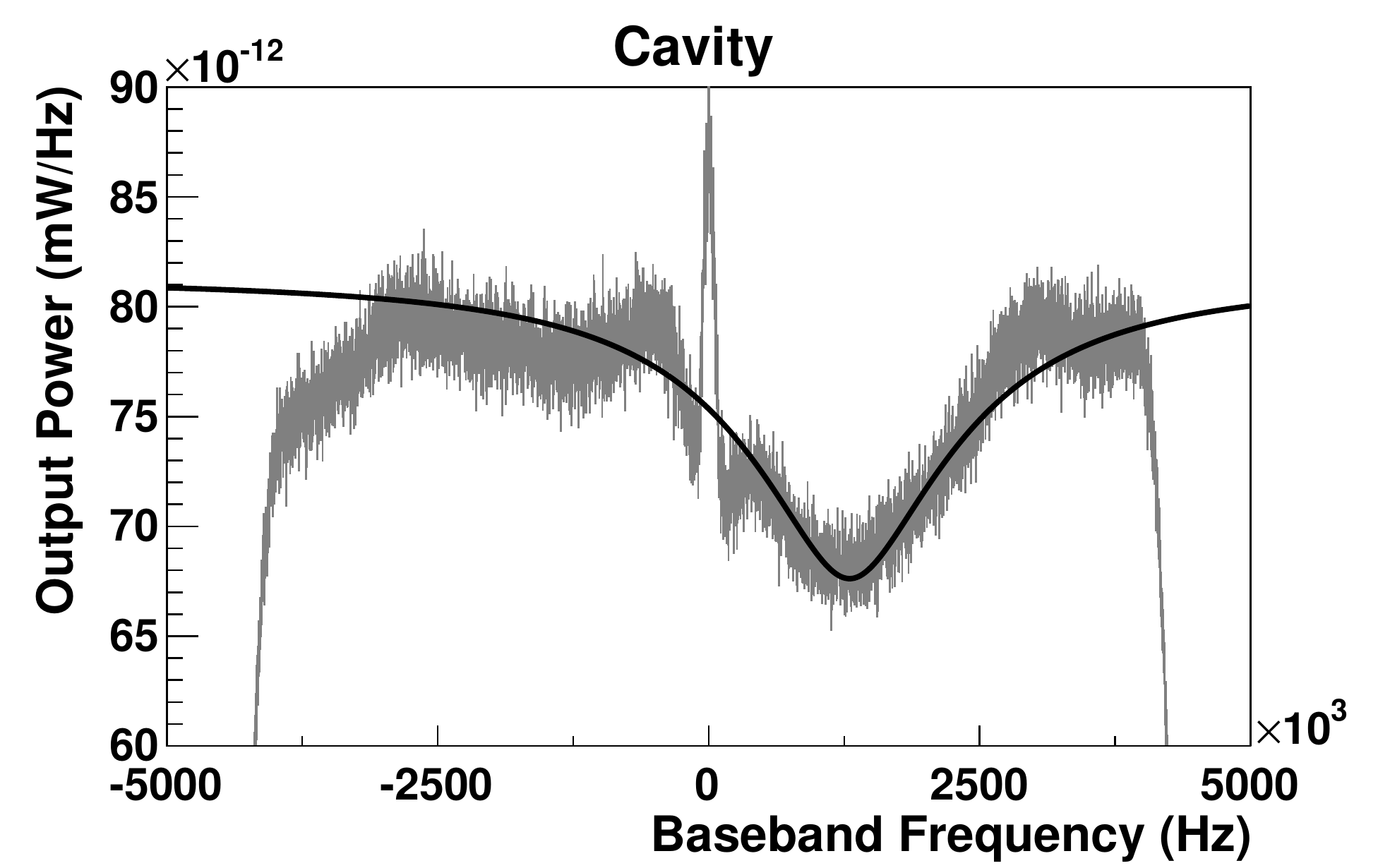}
\end{minipage}
\caption{Upper left panel:  Transmitted power from which the $Q$ of
the TE$_{011}$ cavity is derived.  Upper right:  Power reflected
from the cold TE$_{011}$ cavity tuned to 34.294~GHz.  Lower left:
Power reflected from the cold waveguide to coaxial adapter.  Lower
right:  Wave model (solid line) from equation~\ref{eq:wavemodel}
incorporating the measurements from the other three panels in this
figure.
 \label{fig:wavemodelfit}}
\end{figure}

Squaring the fields and collecting the cross terms, the power at the
input to the HEMT is expected to be
\begin{eqnarray}
T_a + T_{th} + |\Gamma_1|^2T_b + |\Gamma_2|^2T_b|\tau_1|^4 +
2|\Gamma_1|T_ccos(\phi_{s1}+\phi_c) + \nonumber \\
2|\Gamma_2|T_c\tau_1^2cos(\phi_{s12}+\phi_c) + 
2|\Gamma_2||\Gamma_1|T_b\tau_1^2cos(\phi_{s1}+\phi_{s12}).\label{eq:wavemodel}
\end{eqnarray}
The first two terms $T_a+T_{th}$ are equivalent to $T_{sys}$,
estimated to be 27~K in Section~\ref{sec:noisecalibration}. The
parameters $|\Gamma_1|$ and $|\Gamma_2|$ are measured at cryogenic
temperatures in terms of return loss with the Agilent PNA E8364C
network analyzer and are labeled S11 and S22 in
Figure~\ref{fig:wavemodelfit}. Oscillations with frequency 50~MHz
can also be seen on the plots. These features indicate the presence
of well-behaved reflections inside the 6~m round-trip length of
waveguide. The loaded Q of the TE$_{011}$ cavity is measured at 5~K and
is labeled S21 in Figure~\ref{fig:wavemodelfit}. From the S22 of the
cavity on resonance (-12~dB) and the measurement of $Q$ at 10$^4$,
the frequency--dependent $\Gamma_2$ behaves as a Lorenztian
\begin{equation}
|\Gamma_2| = \frac{10.^{-12./20.}}{1+
\left(\frac{2Q(f-f_0)}{f_0}\right)^2}.\nonumber
\end{equation}
Remaining parameters that are unmeasured in the model are the three
phase shifts $\phi_1$, $\phi_{12}$, and $\phi_c$, as well as $T_b$
and $T_c$. The bottom right panel of Figure~\ref{fig:wavemodelfit}
shows an overlay of the model with the data from the receiver with
the 5~K resonant cavity at the input. The free parameters have been
adjusted by hand in the plot to be $T_c$=3.0~K, $T_b$=0.5~K,
$\phi_c$=0, $\phi_{s1}$=0, and $\phi_{s12}$=0. Although the solution
is not a unique one, it is still encouraging that the model can
account for the observations made in this experiment.

\subsection{Data Reduction}
The aim of the data processing is to optimize the signal to noise
ratio
\begin{equation}  
\frac{S}{N} =
\frac{P_{sig}\sqrt{\tau}}{k_BT_{sys}\sqrt{\Delta\nu}}\label{eq:SbyN}
\end{equation}
where $P_{sig}$ is the signal power, $k_B\sigma_T\Delta\nu$ is the
noise power, and $\sigma_T$ is defined in Equation~\ref{eq:dicke}.
The bandwidth of $P_{sig}$ in Equation~\ref{eq:SbyN} is assumed to
be narrower than $\Delta\nu$.  From the expression it is clear that
$P_{sig}$ increases as $\sqrt{\tau}$ and as $1/\sqrt{\Delta\nu}$. It
is therefore beneficial to choose the narrowest possible value of
$\Delta\nu$ while still requiring that it encapsulate the width of
$P_{sig}$.

After selecting the resolution bandwidth, the data are converted
from a raw time series of voltages in $I$ and $Q$ to a power
spectrum in frequency. This is done using a complex Fast Fourier
Transform (FFT) taken from the fftw3~\cite{frigo} libraries.  After
the FFTs there are $N=\tau\Delta\nu$ frequency spectra that are
averaged together.

\subsection{Driven Experiment}
In the case of the two--cavity experiment, the data reduction is a
straightforward search for a monoenergetic signal at the same
frequency as the drive signal.  Also, since the drive signal is
narrow ($<$10~mHz) and is frequency--locked to the receiver chain,
$\Delta\nu$ should be narrow to optimize the signal to noise
ratio~(e.g. \cite{Caspers:2009cj,Betz:2013dza}).

\begin{figure}
\hspace{-0.2in}
\includegraphics[width=3.0in]{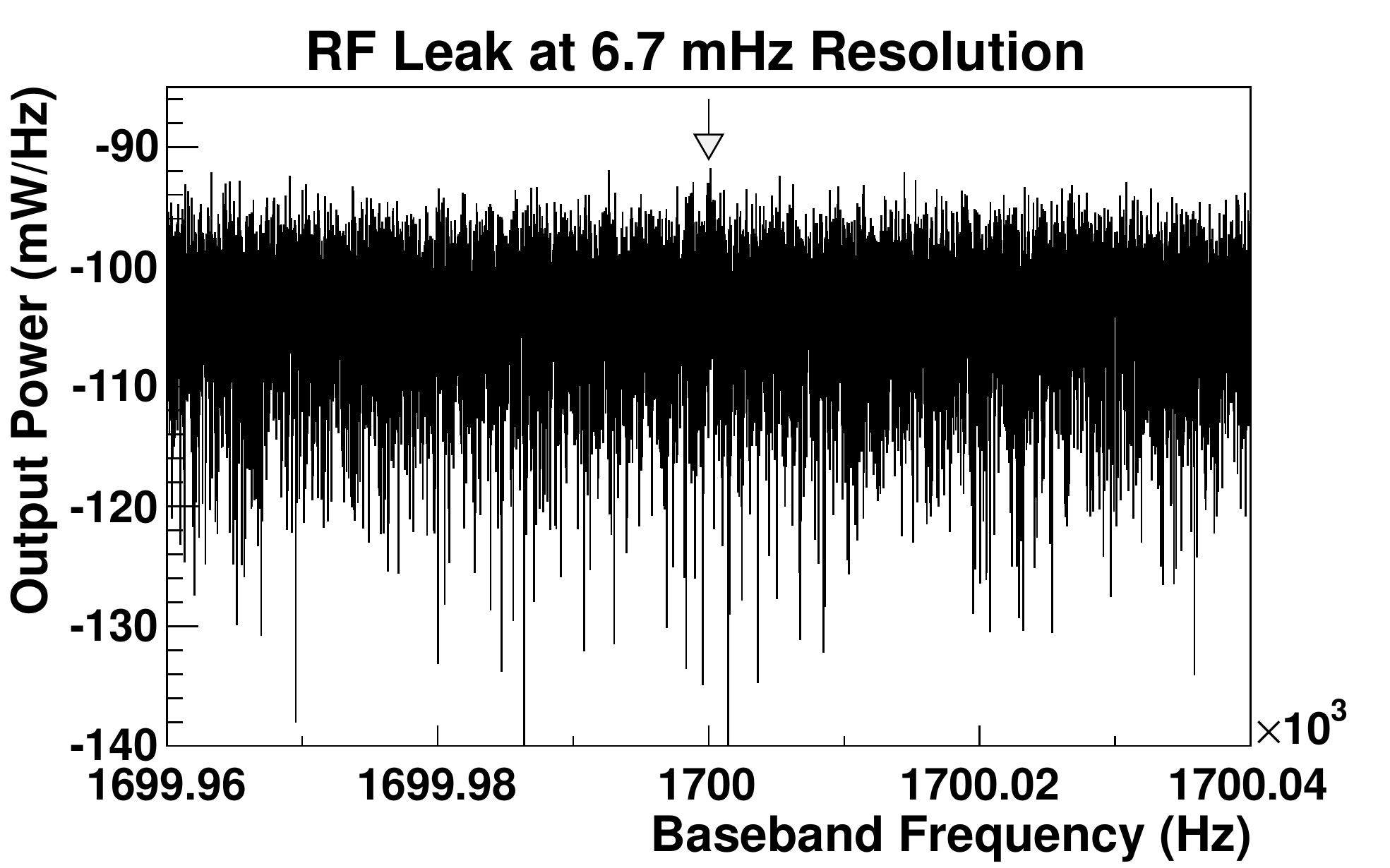}
\includegraphics[width=3.0in]{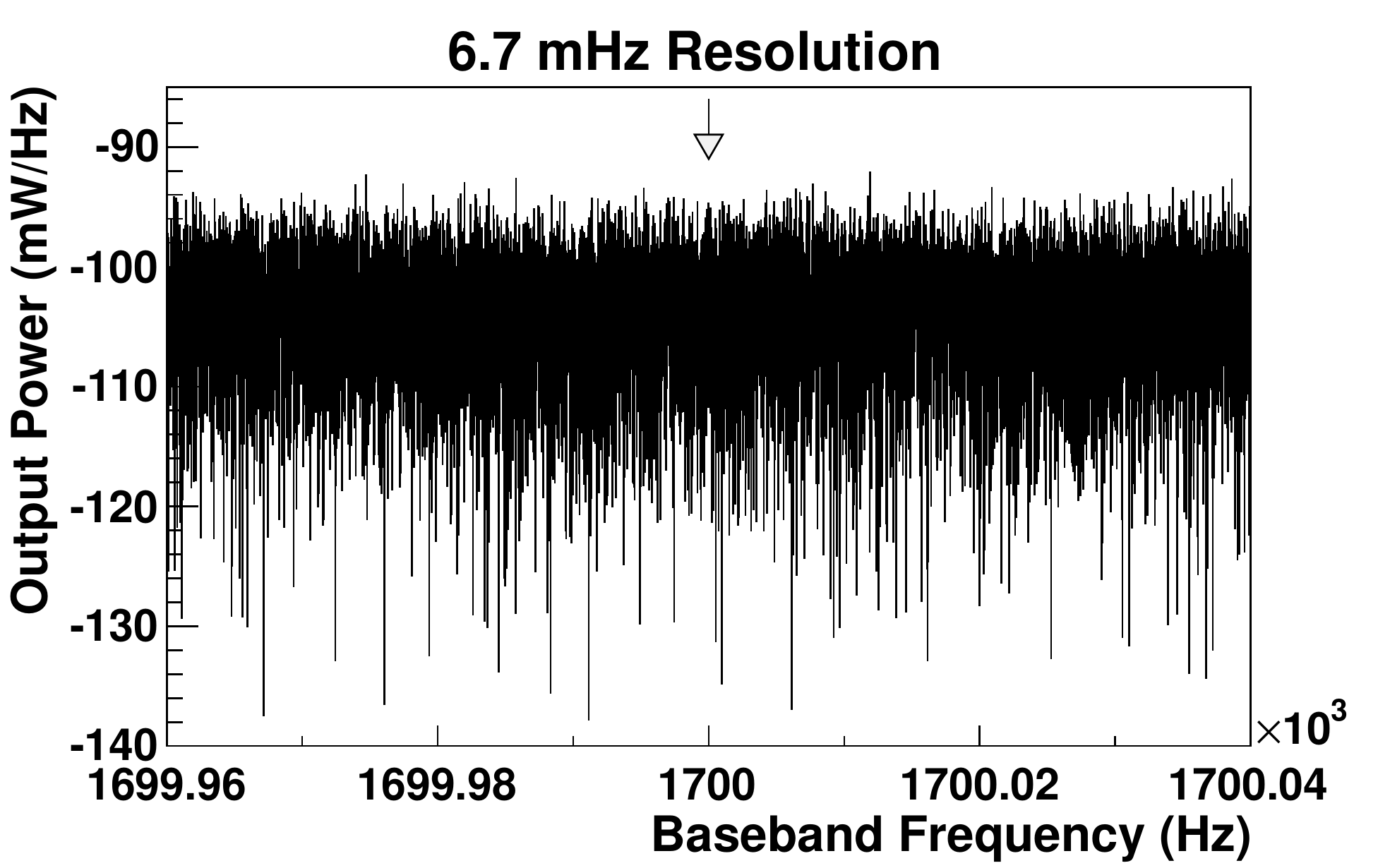}
\caption{The left panel shows the power from the signal cavity in
the presence of an RF leak.  The right panel shows the power
spectrum without the RF leak.  The arrow in each plot marks the
frequency in the baseband where the drive signal should be after
downmixing.\label{fig:LSWShielding}}
\end{figure}

Figure~\ref{fig:LSWShielding} shows an example of a data run with an
RF leak (left panel), and with the RF leak suppressed (right panel).
The expected frequency of the drive signal in the baseband is noted
with an arrow.  In looking at the figure one might notice that the
width of the RF leak appears to span more than one bin.  At first
glance this may be surprising given that the source is narrow,
$\Delta\nu$ is 6.7 mHz, and the system is frequency--locked. However
in this particular measurement the RF leak was found to be related
to a ground loop that included the two cryogenic amplifiers and
their power supplies, which probably means that the power in this RF
leak was not well locked to the rest of the system.  This can
account for the excess width in the signal.

The shielding between the two cavities is demonstrated, also in
Figure~\ref{fig:LSWShielding}.  The power contained in the RF leak
is 10$^{-21}$~Watts.  The signal appears at the expected frequency
in the baseband which demonstrates that the detection and analysis
are working correctly.  With the RF leak suppressed, as in the right
panel, the shielding is estimated from the drive power and
$\sigma_{tot}$ to be $P_{sig}/(k_B\sigma_{tot}\Delta\nu)\geq$223~dB.

\subsection{Listening Experiment}
In the swept search  for halo ALPs with one cavity, the data
analysis begins with the FFT as above in the driven experiment.
Following the FFT are several additional steps that are needed to
account for the interaction of the ALP--driven photons with the
apparatus. These steps are more critical here than they are in the
driven experiment largely because the frequency of the signal is
unknown. Thus it becomes important to characterize each bin for
every cavity frequency before averaging the overlapping scans.  The
steps outlined here are similar to those described
in~\cite{Peng:2000hd}~and~\cite{Daw}.  To date the analysis has been
focused on a search for statistical aberrations that are one bin
wide, using a resolution bandwidth of 34~kHz across a tuning bandwidth
of 600~MHz.  Wider scan ranges are possible but are probably best
approached with an automated experiment.

\subsubsection{Mean subtraction}
The first step in the analysis after the FFT is to extract the raw
fluctuations from around the mean power.  To do this, the mean is
calculated empirically and is subtracted from the power spectrum.
Typically the mean is derived from the data in groups of
$\sim$5~bins at a time; this ensures accuracy without significant
degradation of narrow signals.

\subsubsection{Mismatch}
The impedance mismatch between the cavity and the waveguide acts as
a filter during the measurement.  For the case of a cavity
critically coupled to the waveguide on resonance, half of the power
in the cavity propagates to the waveguide.  Off resonance, less than
half of the power propagates.  The function describing this behavior
in frequency is a Lorentzian curve with the $Q$ of the cavity,
peaking at 0.5 . It is applied to the fluctuations by division,
which worsens the resolution everywhere by at least a factor of 2.  The 
coupling of the TM$_{020}$ cavity to the waveguide has been measured
to be close to critical in~\cite{malagon}.  For the case
of a cavity with some other degree of coupling, the mismatch filter 
is derived similarly but with a different peak value that corresponds 
to the fraction of power transmitted out of the cavity on resonance.

\subsubsection{Noise reduction filter}
The narrow signal that is expected from ALP couplings coexists with
the random thermal and electronic noise characterized by $T_{sys}$.
Therefore the signal to noise ratio is improved by an algorithm that
reduces the random noise power.  The Wiener filter~\cite{wiener} is
one example of a useful and accessible tool for this purpose,
implemented as in~\cite{press}
\begin{equation}
b_i = \mu + \frac{(\sigma^2-\nu^2)}{\sigma^2}(a_i - \mu) \nonumber
\end{equation}
where $a_i$ ($b_i$) is the unfiltered (filtered) fluctuation in bin
$i$, $\sigma$ is the standard deviation in the neighborhood of bin
$i$, $\nu$ is the average value of $\sigma$ near bin $i$, and $\mu$
is the mean of $a_i$ near bin $i$.  The typical granularity of the
filter in this analysis is about 10~bins.

\subsubsection{Axion Lorentzian}
\begin{figure}
\hspace{-1in}
\begin{minipage}{1.0\textwidth}
\includegraphics[width=7.0in]{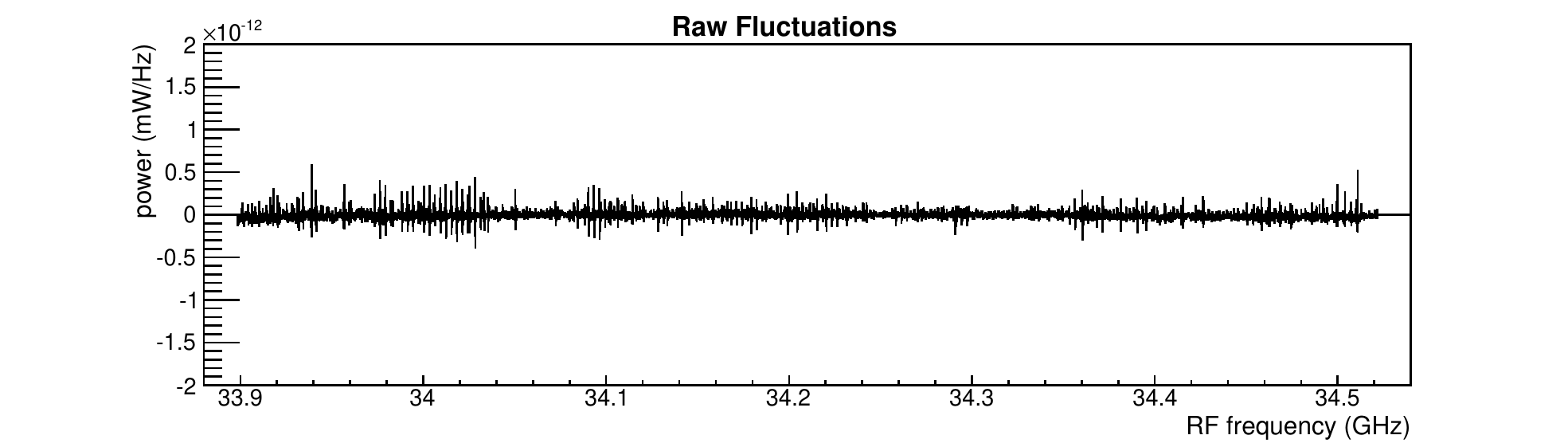}
\includegraphics[width=7.0in]{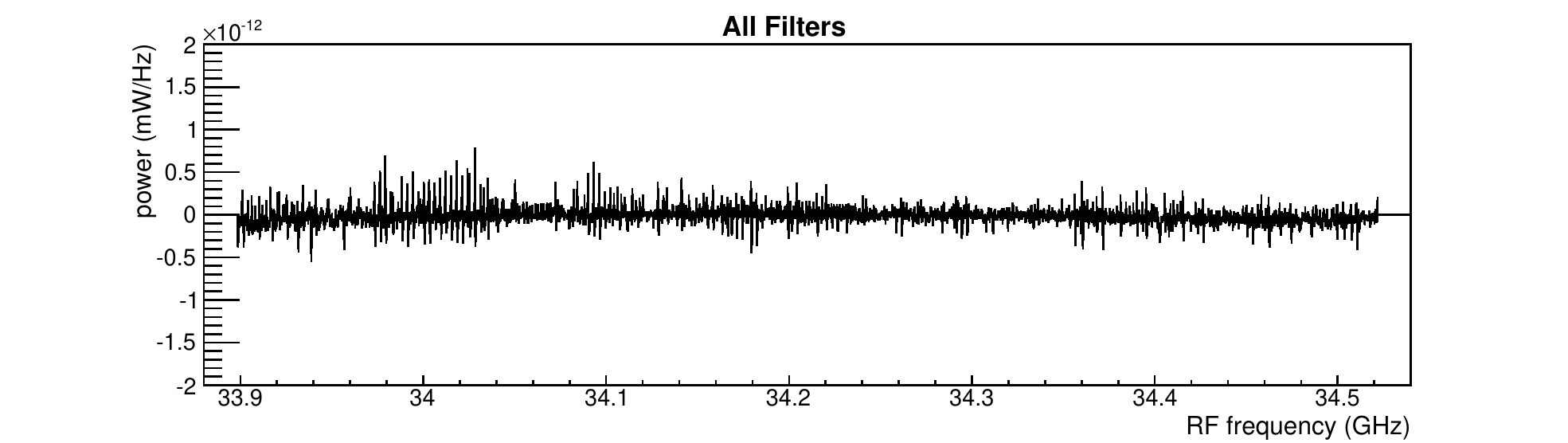}
\end{minipage}
\caption{Two plots showing a sample of raw fluctuations in the data
before (top panel) and after (bottom panel) the data reduction in
the listening experiment, using with one cavity in the TM$_{020}$
mode.  The resolution bandwidth is 34~kHz.\label{fig:datareduction}}
\end{figure}

The last step is to consider the power spectrum of photons that
should come from axion couplings within the volume of the resonant
cavity. The power $P_0$ for the case where the axion mass is equal
to the resonant frequency of the cavity is calculated from the
Lagrangian in~\cite{Sikivie:1983ip,Sikivie:1985yu}. Where the mass
is off resonance, the power spectrum is also derived from the
Lagrangian and follows the Lorentzian shape of the cavity
resonance~\cite{Sikivie:1983ip, Sikivie:1985yu, hagmann}. In
practice this behavior is applied to the data in frequency as a
division by a Lorentzian function that peaks at unity.
Figure~\ref{fig:datareduction} shows a typical data set measured in
the listening mode with the TM$_{020}$ cavity, before and after the
data reduction.

\section{Summary}
The design and nominal behavior of the Yale 34~GHz resonant cavity
experiment has been described.  The noise temperature of the
electronics has been measured.  The observations made with the
apparatus have been compared with expected values, and the data
reduction has been outlined.  The measurements and their derivation
are in accordance both with ideal models and with techniques used in
other experiments~\cite{Sikivie:1983ip,Asztalos:2009yp}.  It is
therefore reasonable to conclude that if an unexpected signal were
to be found in this experiment it should be investigated as a
possible sign of new physics.

\section{Acknowledgments}
The authors are grateful to the United States Office of Naval
Research Directed Energy Program and to Yale University for their
generous financial support.  The authors also wish to thank
Professor Kurt Zilm for the use of the NMR magnet throughout this
experiment.


\bibliography{ymce}

\providecommand{\noopsort}[1]{}\providecommand{\singleletter}[1]{#1}%
\begin{thebibliography}{10}
\expandafter\ifx\csname url\endcsname\relax
  \def\url#1{\texttt{#1}}\fi
\expandafter\ifx\csname urlprefix\endcsname\relax\def\urlprefix{URL }\fi
\expandafter\ifx\csname href\endcsname\relax
  \def\href#1#2{#2} \def\path#1{#1}\fi

\bibitem{Zwicky:1937zza}
F.~Zwicky, {On the Masses of Nebulae and of Clusters of Nebulae}, Astrophys.J.
  86 (1937) 217--246.
\newblock \href {http://dx.doi.org/10.1086/143864} {\path{doi:10.1086/143864}}.

\bibitem{Sofue:2008wt}
Y.~Sofue, M.~Honma, T.~Omodaka, {Unified Rotation Curve of the Galaxy --
  Decomposition into de Vaucouleurs Bulge, Disk, Dark Halo, and the 9-kpc
  Rotation Dip --}, Publ.Astron.Soc.Japan 61 (2009) 227--236.
\newblock \href {http://arxiv.org/abs/0811.0859} {\path{arXiv:0811.0859}}.

\bibitem{Jarosik:2010iu}
N.~Jarosik, C.~Bennett, J.~Dunkley, B.~Gold, M.~Greason, et~al., {Seven-Year
  Wilkinson Microwave Anisotropy Probe (WMAP) Observations: Sky Maps,
  Systematic Errors, and Basic Results}, Astrophys.J.Suppl. 192 (2011) 14.
\newblock \href {http://arxiv.org/abs/1001.4744} {\path{arXiv:1001.4744}},
  \href {http://dx.doi.org/10.1088/0067-0049/192/2/14}
  {\path{doi:10.1088/0067-0049/192/2/14}}.

\bibitem{Ade:2013ktc}
{Tauber, Jan}, \href{http://dx.doi.org/10.1051/0004-6361/201321529}{Planck 2013
  results. i. overview of products and scientific results}, A\&A\href
  {http://dx.doi.org/10.1051/0004-6361/201321529}
  {\path{doi:10.1051/0004-6361/201321529}}.
\newline\urlprefix\url{http://dx.doi.org/10.1051/0004-6361/201321529}

\bibitem{Aalseth:2012if}
C.~Aalseth, et~al., {CoGeNT: A Search for Low-Mass Dark Matter using p-type
  Point Contact Germanium Detectors}, Phys.Rev. D88~(1) (2013) 012002.
\newblock \href {http://arxiv.org/abs/1208.5737} {\path{arXiv:1208.5737}},
  \href {http://dx.doi.org/10.1103/PhysRevD.88.012002}
  {\path{doi:10.1103/PhysRevD.88.012002}}.

\bibitem{Angloher:2011uu}
G.~Angloher, M.~Bauer, I.~Bavykina, A.~Bento, C.~Bucci, et~al., {Results from
  730 kg days of the CRESST-II Dark Matter Search}, Eur.Phys.J. C72 (2012)
  1971.
\newblock \href {http://arxiv.org/abs/1109.0702} {\path{arXiv:1109.0702}},
  \href {http://dx.doi.org/10.1140/epjc/s10052-012-1971-8}
  {\path{doi:10.1140/epjc/s10052-012-1971-8}}.

\bibitem{Bernabei:2008yi}
R.~Bernabei, et~al., {First results from DAMA/LIBRA and the combined results
  with DAMA/NaI}, Eur.Phys.J. C56 (2008) 333--355.
\newblock \href {http://arxiv.org/abs/0804.2741} {\path{arXiv:0804.2741}},
  \href {http://dx.doi.org/10.1140/epjc/s10052-008-0662-y}
  {\path{doi:10.1140/epjc/s10052-008-0662-y}}.

\bibitem{Savage:2008er}
C.~Savage, G.~Gelmini, P.~Gondolo, K.~Freese, {Compatibility of DAMA/LIBRA dark
  matter detection with other searches}, JCAP 0904 (2009) 010.
\newblock \href {http://arxiv.org/abs/0808.3607} {\path{arXiv:0808.3607}},
  \href {http://dx.doi.org/10.1088/1475-7516/2009/04/010}
  {\path{doi:10.1088/1475-7516/2009/04/010}}.

\bibitem{Agnese:2013rvf}
R.~Agnese, et~al., {Silicon Detector Dark Matter Results from the Final
  Exposure of CDMS II}, Phys.Rev.Lett. 111 (2013) 251301.
\newblock \href {http://arxiv.org/abs/1304.4279} {\path{arXiv:1304.4279}},
  \href {http://dx.doi.org/10.1103/PhysRevLett.111.251301}
  {\path{doi:10.1103/PhysRevLett.111.251301}}.

\bibitem{Aprile:2013doa}
E.~Aprile, et~al., {Limits on spin-dependent WIMP-nucleon cross sections from
  225 live days of XENON100 data}, Phys.Rev.Lett. 111~(2) (2013) 021301.
\newblock \href {http://arxiv.org/abs/1301.6620} {\path{arXiv:1301.6620}},
  \href {http://dx.doi.org/10.1103/PhysRevLett.111.021301}
  {\path{doi:10.1103/PhysRevLett.111.021301}}.

\bibitem{Angle:2011th}
J.~Angle, et~al., {A search for light dark matter in XENON10 data},
  Phys.Rev.Lett. 107 (2011) 051301.
\newblock \href {http://arxiv.org/abs/1104.3088} {\path{arXiv:1104.3088}},
  \href {http://dx.doi.org/10.1103/PhysRevLett.110.249901,
  10.1103/PhysRevLett.107.051301} {\path{doi:10.1103/PhysRevLett.110.249901,
  10.1103/PhysRevLett.107.051301}}.

\bibitem{Armengaud:2011cy}
E.~Armengaud, et~al., {Final results of the EDELWEISS-II WIMP search using a
  4-kg array of cryogenic germanium detectors with interleaved electrodes},
  Phys.Lett. B702 (2011) 329--335.
\newblock \href {http://arxiv.org/abs/1103.4070} {\path{arXiv:1103.4070}},
  \href {http://dx.doi.org/10.1016/j.physletb.2011.07.034}
  {\path{doi:10.1016/j.physletb.2011.07.034}}.

\bibitem{Armengaud:2012pfa}
E.~Armengaud, et~al., {A search for low-mass WIMPs with EDELWEISS-II
  heat-and-ionization detectors}, Phys.Rev. D86 (2012) 051701.
\newblock \href {http://arxiv.org/abs/1207.1815} {\path{arXiv:1207.1815}},
  \href {http://dx.doi.org/10.1103/PhysRevD.86.051701}
  {\path{doi:10.1103/PhysRevD.86.051701}}.

\bibitem{Akimov:2011tj}
D.~Y. Akimov, H.~Araujo, E.~Barnes, V.~Belov, A.~Bewick, et~al., {WIMP-nucleon
  cross-section results from the second science run of ZEPLIN-III}, Phys.Lett.
  B709 (2012) 14--20.
\newblock \href {http://arxiv.org/abs/1110.4769} {\path{arXiv:1110.4769}},
  \href {http://dx.doi.org/10.1016/j.physletb.2012.01.064}
  {\path{doi:10.1016/j.physletb.2012.01.064}}.

\bibitem{Akerib:2013tjd}
D.~Akerib, et~al., {First results from the LUX dark matter experiment at the
  Sanford Underground Research Facility}, Phys.Rev.Lett. 112 (2014) 091303.
\newblock \href {http://arxiv.org/abs/1310.8214} {\path{arXiv:1310.8214}},
  \href {http://dx.doi.org/10.1103/PhysRevLett.112.091303}
  {\path{doi:10.1103/PhysRevLett.112.091303}}.

\bibitem{ATLAS:2012ky}
G.~Aad, et~al., {Search for dark matter candidates and large extra dimensions
  in events with a jet and missing transverse momentum with the ATLAS
  detector}, JHEP 1304 (2013) 075.
\newblock \href {http://arxiv.org/abs/1210.4491} {\path{arXiv:1210.4491}},
  \href {http://dx.doi.org/10.1007/JHEP04(2013)075}
  {\path{doi:10.1007/JHEP04(2013)075}}.

\bibitem{Peccei:1977hh}
R.~Peccei, H.~R. Quinn, {CP Conservation in the Presence of Instantons},
  Phys.Rev.Lett. 38 (1977) 1440--1443.
\newblock \href {http://dx.doi.org/10.1103/PhysRevLett.38.1440}
  {\path{doi:10.1103/PhysRevLett.38.1440}}.

\bibitem{Weinberg:1977ma}
S.~Weinberg, {A New Light Boson?}, Phys.Rev.Lett. 40 (1978) 223--226.
\newblock \href {http://dx.doi.org/10.1103/PhysRevLett.40.223}
  {\path{doi:10.1103/PhysRevLett.40.223}}.

\bibitem{Wilczek:1977pj}
F.~Wilczek, {Problem of Strong p and t Invariance in the Presence of
  Instantons}, Phys.Rev.Lett. 40 (1978) 279--282.
\newblock \href {http://dx.doi.org/10.1103/PhysRevLett.40.279}
  {\path{doi:10.1103/PhysRevLett.40.279}}.

\bibitem{Jaeckel:2007ch}
J.~Jaeckel, A.~Ringwald, {A Cavity Experiment to Search for Hidden Sector
  Photons}, Phys.Lett. B659 (2008) 509--514.
\newblock \href {http://arxiv.org/abs/0707.2063} {\path{arXiv:0707.2063}},
  \href {http://dx.doi.org/10.1016/j.physletb.2007.11.071}
  {\path{doi:10.1016/j.physletb.2007.11.071}}.

\bibitem{Arias:2012mb}
P.~Arias, D.~Cadamuro, M.~Goodsell, J.~Jaeckel, J.~Redondo, et~al., {WISPy Cold
  Dark Matter, DESY-11-226, MPP-2011-140, CERN-PH-TH-2011-323, IPPP-11-80,
  DCPT-11-160}\href {http://arxiv.org/abs/1201.5902} {\path{arXiv:1201.5902}}.

\bibitem{Nelson:2011sf}
A.~E. Nelson, J.~Scholtz, {Dark Light, Dark Matter and the Misalignment
  Mechanism}, Phys.Rev. D84 (2011) 103501.
\newblock \href {http://arxiv.org/abs/1105.2812} {\path{arXiv:1105.2812}},
  \href {http://dx.doi.org/10.1103/PhysRevD.84.103501}
  {\path{doi:10.1103/PhysRevD.84.103501}}.

\bibitem{Sikivie:1983ip}
P.~Sikivie, {Experimental Tests of the Invisible Axion}, Phys.Rev.Lett. 51
  (1983) 1415--1417.
\newblock \href {http://dx.doi.org/10.1103/PhysRevLett.51.1415}
  {\path{doi:10.1103/PhysRevLett.51.1415}}.

\bibitem{Sikivie:1985yu}
P.~Sikivie, {Detection rates for 'invisible' axion searches}, Phys. Rev. D32
  (1985) 2988.
\newblock \href {http://dx.doi.org/10.1103/PhysRevD.36.974,
  10.1103/PhysRevD.32.2988} {\path{doi:10.1103/PhysRevD.36.974,
  10.1103/PhysRevD.32.2988}}.

\bibitem{Peng:2000hd}
H.~Peng, S.~J. Asztalos, E.~Daw, N.~Golubev, C.~Hagmann, et~al., {Cryogenic
  cavity detector for a large scale cold dark-matter axion search},
  Nucl.Instrum.Meth. A444 (2000) 569--583.
\newblock \href {http://dx.doi.org/10.1016/S0168-9002(99)00971-7}
  {\path{doi:10.1016/S0168-9002(99)00971-7}}.

\bibitem{bradley}
R.~Bradley, J.~Clarke, D.~Kinion, L.~Rosenberg, K.~van Bibber, et~al.,
  {Microwave cavity searches for dark-matter axions}, Rev. Mod. Phys. 75 (2003)
  777--817.
\newblock \href {http://dx.doi.org/10.1103/RevModPhys.75.777}
  {\path{doi:10.1103/RevModPhys.75.777}}.

\bibitem{Asztalos:2009yp}
S.~Asztalos, et~al., {A SQUID-based microwave cavity search for dark-matter
  axions}, Phys.Rev.Lett. 104 (2010) 041301.
\newblock \href {http://arxiv.org/abs/0910.5914} {\path{arXiv:0910.5914}},
  \href {http://dx.doi.org/10.1103/PhysRevLett.104.041301}
  {\path{doi:10.1103/PhysRevLett.104.041301}}.

\bibitem{malagon}
A.~T. Malagon, {Search for 140 $\mu$eV Pseudoscalar and Vector Dark Matter
  Using Microwave Cavities}, Ph.D. Thesis, Yale University (2014).

\bibitem{weinreb}
S.~Weinreb, M.~W. Pospieszalski, R.~Norrod, {Cryogenic HEMT Low-Noise Receivers
  for 1.3 to 43 GHz Range}, Microwave Symposium Digest, 1988., IEEE MTT-S
  International 2 (1988) 945--948.
\newblock \href {http://dx.doi.org/10.1109/MWSYM.1988.22187}
  {\path{doi:10.1109/MWSYM.1988.22187}}.

\bibitem{Dicke}
R.~H. Dicke, Rev. Sci Instrum. 17 (1946) 268.

\bibitem{nist260-46}
J.~G. Hust, P.~J. Giarratano, Standard Reference Materials: Thermal
  Conductivity and Electrical Resistivity. Standard Reference Materials:
  Austenitic Stainless Steel, SRMs 735 and 798, From 4 to 1200 K., Nat. Bur.
  Stand. (U.S.) Spec. Publ. 260-46, Washington, D.C., 1975.

\bibitem{reid2008}
M.~Reid, Low-noise systems in the Deep Space Network, Wiley, Hoboken, N.J,
  2008.

\bibitem{Daw:1997}
E.~Daw, R.~F. Bradley, {Effect of High Magnetic Fields on the Noise Temperature
  of a Heterostructure Field-Effect Transistor Low-Noise Amplifier},
  J.Appl.Phys. 82 (1997) 1925.
\newblock \href {http://dx.doi.org/10.1063/1.366000}
  {\path{doi:10.1063/1.366000}}.

\bibitem{condon}
J.~J. Condon, S.~M. Ransom,
  \href{www.cv.nrao.edu/course/astr534/ERA.shtml}{{Essential radio astronomy}}.
\newline\urlprefix\url{www.cv.nrao.edu/course/astr534/ERA.shtml}

\bibitem{Meys}
R.~P. Meys, {A wave approach to the noise properties of linear microwave
  devices}, IEEE Trans. on Microwave Theory and Tech. 26~(1) (1978) 34--37.

\bibitem{wedge}
S.~W. Wedge, D.~B. Rutledge, {Wave Techniques for Noise Modeling and
  Measurement}, IEEE Trans. on Microwave Theory and Tech. 40~(11) (1992)
  2004--2013.

\bibitem{frigo}
M.~Frigo, S.~G. Johnson, {The design and implementation of FFTW3}, Proc. IEEE
  93 (2005) 216--231.
\newblock \href {http://dx.doi.org/10.1109/JPROC.2004.840301}
  {\path{doi:10.1109/JPROC.2004.840301}}.

\bibitem{Caspers:2009cj}
F.~Caspers, J.~Jaeckel, A.~Ringwald, {Feasibility, engineering aspects and
  physics reach of microwave cavity experiments searching for hidden photons
  and axions}, JINST 4 (2009) P11013.
\newblock \href {http://arxiv.org/abs/0908.0759} {\path{arXiv:0908.0759}},
  \href {http://dx.doi.org/10.1088/1748-0221/4/11/P11013}
  {\path{doi:10.1088/1748-0221/4/11/P11013}}.

\bibitem{Betz:2013dza}
M.~Betz, F.~Caspers, M.~Gasior, M.~Thumm, S.~Rieger, {First results of the CERN
  Resonant Weakly Interacting sub-eV Particle Search (CROWS)}, Phys.Rev.
  D88~(7) (2013) 075014.
\newblock \href {http://arxiv.org/abs/1310.8098} {\path{arXiv:1310.8098}},
  \href {http://dx.doi.org/10.1103/PhysRevD.88.075014}
  {\path{doi:10.1103/PhysRevD.88.075014}}.

\bibitem{Daw}
E.~J. Daw, {A search for halo axions}, Ph.D. Thesis, Massachusetts Institute of
  Technology (1998).

\bibitem{wiener}
N.~Wiener, {The interpolation, extrapolation and smoothing of stationary time
  series}, New York: Wiley, 1949.

\bibitem{press}
W.~H. Press, S.~A. Teulosky, W.~T. Vetterling, B.~P. Flannery, {Numerical
  Recipes 3rd Edition}, Cambridge University Press, 2007.

\bibitem{hagmann}
C.~A. Hagmann, {A search for cosmic axions}, Ph.D. Thesis, University of
  Florida (1990).

\end{thebibliography}

\end{document}